\documentclass[prd,preprintnumbers,aps,preprint,amsmath,amssymb,superscriptaddress,floatfix,nofootinbib,unsortedaddress]{revtex4}
\pdfoutput=1

\usepackage{amssymb, bm, amsmath, graphicx, physymb}
\usepackage[usenames,dvipsnames]{color}
\usepackage{slashed}
\usepackage[utf8]{inputenc}
\usepackage{enumerate}
\usepackage{amsfonts}
\usepackage{mathtools}
\usepackage{latexsym}
\usepackage{epstopdf} % COMPILE EPS (AND PDF) FIGURES USING PDFLATEX!!!
\usepackage{bbm}
\usepackage{soul}
\usepackage[papersize={8.5in,11in}]{geometry} % Fixes annoying href error
\usepackage[colorlinks=true,linktocpage=true,linkcolor=blue,citecolor=blue,hyperfootnotes=false]{hyperref}

\definecolor{carnelian}{rgb}{0.7, 0.11, 0.11}
\definecolor{ferngreen}{rgb}{0.31, 0.47, 0.26}

\newcommand{\T}{T}
\newcommand{\PrpXY}{\bot}
\newcommand{\Tvec}[1]{\vec{#1}_\T}
\newcommand{\tvec}[1]{\vec{#1}_\PrpXY}

\newcommand{\ord}[1]{\mathcal{O} \left( #1 \right)}

\newcommand{\eq}[1]{Eq.~(\ref{#1})}

\begin{document}
%
%%%%%%%%%%%%%%%%%%%%%%%%%%%%%%%%%%%%%%%%%%%%%%%%%%%%%%%%%%%%%%%%%%%%%%%%%%%
%%%%%%%%%%%%%%%%%%%%%%%%%%%%%%%%%%%%%%%%%%%%%%%%%%%%%%%%%%%%%%%%%%%%%%%%%%%

%%%%%%%%%%%%%%%%%%%%%%%%%%%%%%%%%%%%%%%%%%%%%%%%%%%%%%%%%%%%%%%%%%%%%%%%%%%
%
\title{Jet Drift and Collective Flow in Heavy-Ion Collisions}
%
%%%%%%%%%%%%%%%%%%%%%%%%%%%%%%%%%%%%%%%%%%%%%%%%%%%%%%%%%%%%%%%%%%%%%%%%%%%

%--------------------------------------------------------------------------
\author{Logan Antiporda}
\email[Email: ]{lma2000@nmsu.edu}
\affiliation{Department of Physics, New Mexico State University, Las Cruces, NM 88003, USA}
%--------------------------------------------------------------------------
\author{Joseph Bahder}
\email[Email: ]{jbahder@nmsu.edu}
\affiliation{Department of Physics, New Mexico State University, Las Cruces, NM 88003, USA}
%--------------------------------------------------------------------------
\author{Hasan Rahman}
\email[Email: ]{hrrahman@nmsu.edu}
\affiliation{Department of Physics, New Mexico State University, Las Cruces, NM 88003, USA}
%--------------------------------------------------------------------------
\author{Matthew D. Sievert}
\email[Email: ]{msievert@nmsu.edu}
\affiliation{Department of Physics, New Mexico State University, Las Cruces, NM 88003, USA}
%--------------------------------------------------------------------------

%%%%%%%%%%%%%%%%%%%%%%%%%%%%%%%%%%%%%%%%%%%%%%%%%%%%%%%%%%%%%%%%%%%%%%%%%%

\begin{abstract}
    We study the tomographic applications of a new phenomenon we dub ``jet drift'' -- the deflection of high-energy particles and jets toward the direction of a flowing medium -- to the quark-gluon plasma produced in heavy-ion collisions.  While the physics of jet drift is quite general, for specificity we consider the case of photon-jet production at mid-rapidity.  Beginning with the simplest possible model, a large slab of uniformly flowing plasma, we systematically introduce the geometrical elements of a heavy-ion collision in a simple optical Glauber model.  We find that the moving medium causes the jet to drift in the direction of the flow, bending its trajectory and leaving detailed signatures of the flow pattern in the distribution of $\gamma + \: \mathrm{jet}$ acoplanarities.  In the elliptical geometries produced in non-central collisions, this drift effect leads to a strong geometry coupling which persists despite the addition of event-by-event fluctuations in the jet production point, impact parameter, and acoplanarity.  We propose a new observable to measure the jet drift effect through the correlation of $\gamma + \: \mathrm{jet}$ acoplanarities with the elliptic flow of soft particles.  Preliminary estimates suggest this correlation may be studied at sPHENIX and the LHC.
\end{abstract}

%%%%%%%%%%%%%%%%%%%%%%%%%%%%%%%%%%%%%%%%%%%%%%%%%%%%%%%%%%%%%%%%%%%%%%%%%%

\date{\today}
\maketitle
\tableofcontents
\newpage

%%%%%%%%%%%%%%%%%%%%%%%%%%%%%%%%%%%%%%%%%%%%%%%%%%%%%%%%%%%%%%%%%%%
%
\section{Introduction}
\label{sec:intro}
%
%%%%%%%%%%%%%%%%%%%%%%%%%%%%%%%%%%%%%%%%%%%%%%%%%%%%%%%%%%%%%%%%%%%

In ultrarelativistic collisions of heavy ions, hard processes like jet production provide the natural short-distance tomographic probes of the produced quark-gluon plasma.  Jets and their associated pattern of soft gluon radiation carry detailed interferometric information about the plasma through the Landau-Pomeranchuk-Migdal effect \cite{Landau:1953um, Migdal:1956tc}, which describes the medium-induced modification of the radiation spectrum.  The principle of jet tomography has been an enduring goal of the heavy-ion programs at RHIC and the LHC \cite{Vitev:2002pf, JET:2013cls, Cao:2020wlm}, but in practice, extracting the medium information carried by the jets is challenging.  Jet quenching, the suppression of jet yields in heavy-ion collisions, is unambiguously established in bread-and-butter measurements of the nuclear modification factor, $R_{AA}$, yet is so robust an observable that it can be described by a wide range of divergent models and approximations \cite{PHENIX:2001hpc, ATLAS:2018gwx}.  More differential observables, such as jet shapes \cite{CMS:2013lhm, CMS:2019btm}, groomed jets \cite{Larkoski:2014wba}, and jet substructure \cite{Chien:2016led} can provide better model discrimination, but rapidly become challenging experimentally due to limited statistics and background subtraction.  Even the simplest consequence of medium modification, the Gaussian broadening of the jet distribution relative to its initial direction, has proved difficult to distinguish from the Sudakov radiative broadening in vacuum \cite{Sudakov:1954sw, ALICE:2015mdb}.

\begin{figure}
\begin{centering}
\includegraphics[width=0.48\textwidth]{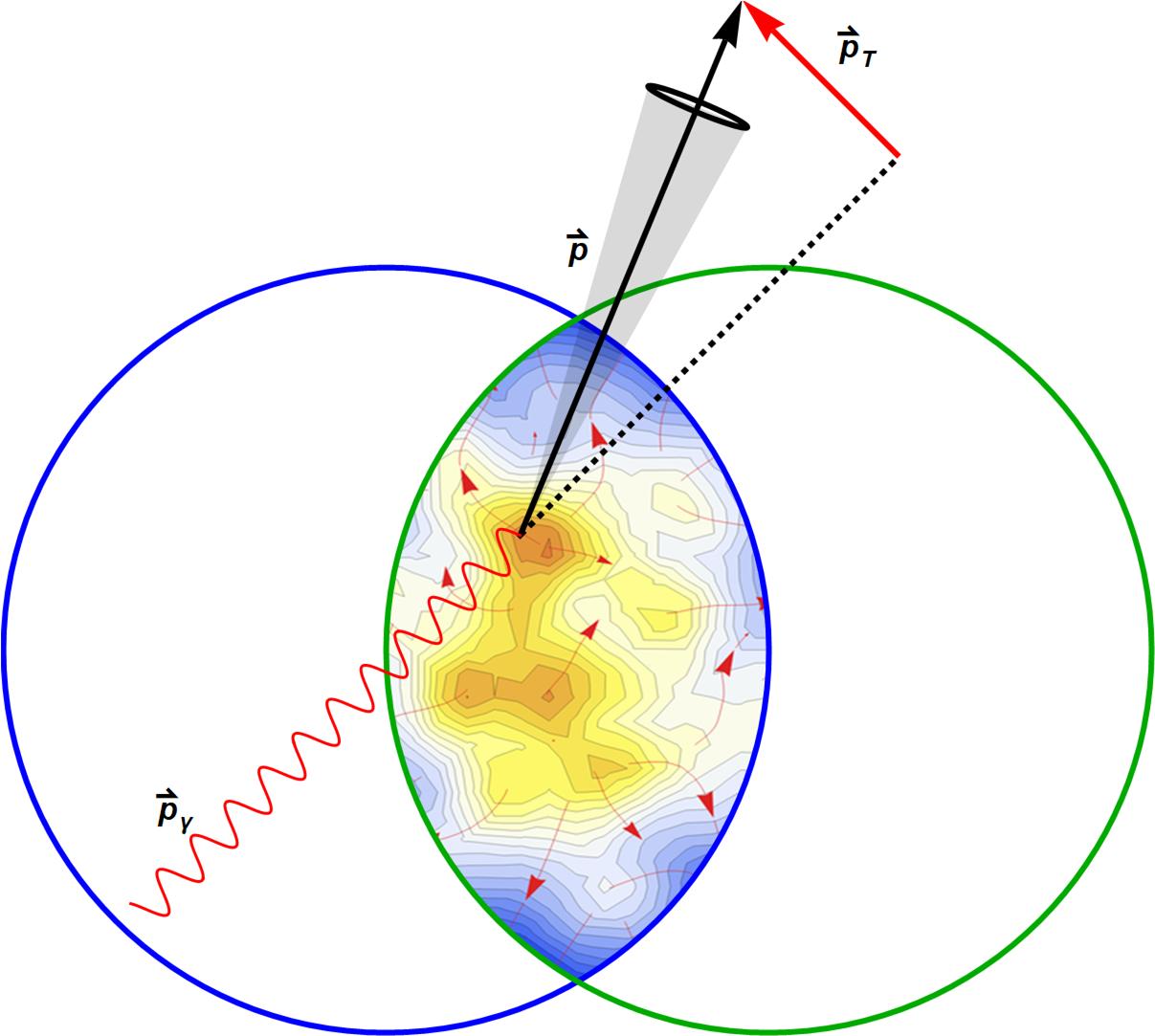}
\includegraphics[width=0.48\textwidth]{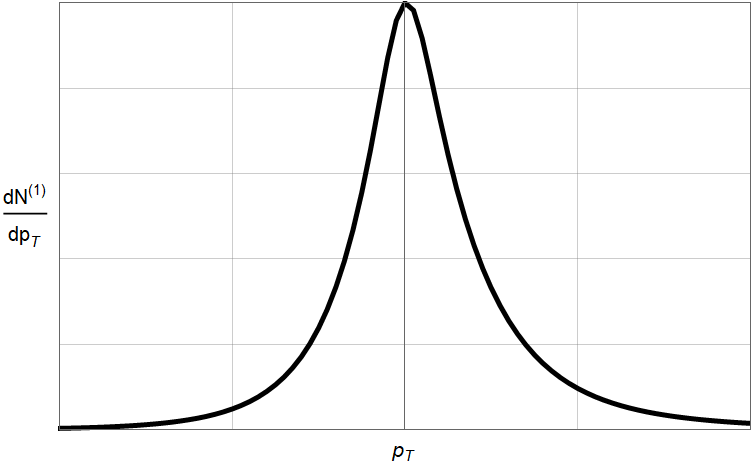}
\caption{Left:  Illustration of a photon + jet pair produced in an off-central heavy-ion collision.  Even if the jet is initially back-to-back with the photon, it picks up some momentum transverse to the photon axis during its interaction with the medium.
Right: Sketch of the acoplanarity distribution for a simulated heavy-ion collision.  The skewness clearly seen in the tail of the distribution is responsible for biasing the jet deflection to the left or the right.
\label{f:process}
}
\end{centering}
\end{figure}

Part of the reason the observation and interpretation of jet modification in heavy-ion collisions has proved challenging is the large number of background effects, both in the medium and in vacuum, which can contribute to standard jet observables like acoplanarities and quenching.  These observables, and the physics they are designed to probe, are symmetric measures of the jet distribution -- its energy loss, its RMS angular deflection, and so on.  They measure the magnitude of the jet scattering in the medium, but are isotropic, making no distinction between scattering in one direction versus another.  For example, consider a hard scattering event in the medium which produces a photon with final momentum $p_\gamma$ and jet with momentum $p$ according to the doubly-differential distribution
\begin{align}
    E_\gamma E \frac{dN}{d^3 p_\gamma \, d^3 p} &\equiv \frac{1}{\sigma_\mathrm{tot}^{\gamma + \mathrm{jet}}} \: E_\gamma E \frac{d\sigma}{d^3 p_\gamma \, d^3 p} 
\end{align}
as depicted in Fig.~\ref{f:process} left.  If the photon and jet are initially produced back-to-back by the hard scattering (as in perturbative QCD at leading order), then in the absence of modification by the medium (denoted as ``$0^\mathrm{th}$ order in the opacity''), the final-state distribution is given by
\begin{align}   \label{e:gamjet0}
    E_\gamma E \, \frac{dN^{(0)}}{d^3 p_\gamma \, d^3 p} &= 
    E_\gamma \frac{dN^{(0)}}{d^3 p_\gamma} \:
    E \: \delta^3 (\vec{p} + \vec{p}_\gamma) \: .
\end{align}
After including the final-state rescatterings in the medium through the opacity series, the exchange of momentum $\vec{p}_T$ \textit{transverse to the photon axis} then leads to a broadening of the initial delta function into a continuous distribution of finite width\footnote{Note that we are using $\vec{p}_T$ to refer to momentum in the plane \textit{transverse to the photon} momentum $\vec{p}_\gamma$, not necessarily \textit{transverse to the beam axis}.  (See Fig.~\ref{f:coords} illustrating the kinematics.)  For jets produced at mid-rapidity, the momentum transverse to the beam axis (also commonly denoted $p_T$) is synonymous with the jet energy.  To avoid confusion, we will refer to the jet kinematics explicitly in terms of its energy.}.

The acoplanarity $\Delta \theta \approx p_T / E$ has mean-squared width
\begin{align}   \label{e:width}
    \langle (\Delta \theta)^2 \rangle =
    \frac{\langle p_T^2 \rangle}{E^2} =
    \int d^3 p \, d^3 p_\gamma \: \left( \frac{p_T^2}{E^2} \right)
    \: \frac{dN}{d^3 p_\gamma \, d^3 p}
\end{align}
reflecting the total momentum broadening induced by the medium.  The width of the acoplanarity distribution is thus an even moment of the jet transverse momentum distribution and depends only on scalar quantities like $p_T^2$ in the plane transverse to the photon axis.  While these even moments are indeed sensitive to modifications induced by the medium, they also receive contributions from other sources such as vacuum Sudakov radiation \cite{Sudakov:1954sw} which are also symmetric in nature \cite{Clayton:2021tbd}.

%--------------------------------------------------------------------------
%--------------------------------------------------------------------------

More recently, a new class of jet observables have been proposed which are \textit{antisymmetric} measures of the jet distribution \cite{He:2020iow, Sadofyev:2021ohn}.  Odd moments like the average acoplanarity $\langle \Delta \theta \rangle$ are zero in the absence of a preferred direction which can skew the distribution asymmetrically, as shown in Fig.~\ref{f:process}.  Expressing these parity-odd quantities as vectors, e.g.
\begin{align}   \label{e:meanvec1}
    \langle \Delta \vec\theta \rangle =
    \int d^3 p \, d^3 p_\gamma \: \left( \frac{\vec{p}_T}{E} \right)
    \: \frac{dN}{d^3 p_\gamma \, d^3 p}
\end{align}
makes manifest the need for a vector direction entering the jet distribution to produce a nonzero result.  Absent an external vector quantity such as a fixed polarization, these antisymmetric observables \textit{cannot} arise from the isotropic mechanisms like vacuum Sudakov radiation which complicate the interpretation of the usual symmetric observables.  Thus asymmetries in observables like the acoplanarity can \textit{only} come from the injection of preferred vector directions from the medium, removing the backgrounds which complicate the study of jet tomography.  Two sources of these asymmetries have been proposed, arising from either the vector direction of gradients \cite{He:2020iow,Sadofyev:2021ohn, Mitkin:2021tbd} (in temperature, density, etc.) or velocities \cite{Sadofyev:2021ohn}.  We also note that, alternatively to a description in perturbative QCD, one can perform a similar analysis in strongly-coupled holographic models and directly compare the drift effects predicted in the two formalisms \cite{Lekaveckas:2013lha, Rajagopal:2015roa, Sadofyev:2015hxa, Li:2016bbh, Reiten:2019fta, Casalderrey-Solana:2014bpa, Brewer:2017fqy}.

In this paper, we apply the theoretical framework for sub-eikonal jet-flow coupling constructed in Ref.~\cite{Sadofyev:2021ohn} to study asymmetries in the $\gamma + \: \mathrm{jet}$ acoplanarity distribution for a range of simple medium models.  We focus on the simplest theoretical case of photon + jet acoplanarities, with the trigger photon providing an unmodified signal of the initial jet direction and the odd moments of the resulting acoplanarity distribution reflecting the net deflection of jets relative to that axis.  

By starting with the simplest possible implementation -- a constant slab of flowing plasma -- and then systematically incorporating the effects of finite geometries and event-by-event fluctuations, we elucidate the elementary physics of jet-flow coupling and its consequences for heavy-ion collisions.  We find that the fundamental effect of the medium velocity is ``jet drift,'' a drag effect which pulls the jet distribution preferentially in the direction of the velocity, and that when applied to the elliptical geometries produced in off-central heavy-ion collisions, this results in a significant geometry-driven coupling to the event plane angle.  We propose a new observable to measure the jet drift effect, the correlation of the $\gamma + \: \mathrm{jet}$ acoplanarity with the elliptic flow in the soft sector, and we substantiate this proposal with event-by-event simulations verifying the robustness of such a correlation.

The rest of this paper is organized as follows.  Sec.~\ref{sec:slab} is dedicated to identifying the elementary physics of jet drift in a moving medium, beginning with a review in Sec.~\ref{sec:moments} of the key results from Ref.~\cite{Sadofyev:2021ohn}.  Evaluating these quantities for the simplest case of a constant slab of flowing plasma, we show in Sec.~\ref{sec:slabmoments} that the asymmetric moments of the acoplanarity distribution carry detailed signatures of the magnitude and direction of the flow.  Further analysis of jets' trajectories through the constant slab is performed in Appendix~\ref{sec:trajectory}.  Sec.~\ref{sec:geometry} builds upon these results by systematically introducing the essential geometric elements of a heavy-ion collision in a simple model, the basics of which we introduce in Sec.~\ref{sec:model}.  Beginning with the mean-field elliptical geometry in Sec.~\ref{sec:staticellipse}, we implement event-by-event fluctuations in the event geometry, jet production point, and acoplanarity in Sec.~\ref{sec:fluct}.  This procedure clearly reveals a strong coupling of the acoplanarity to the event plane, which we quantify through the correlation to the ellipticicity in the initial state and the elliptic flow in the final state.  We summarize the main conclusions in Sec.~\ref{sec:concl} and outline the next steps to generalize these results to other jet processes and state-of-the-art hydrodynamic backgrounds. We also include a study of the model dependence of our main results in Appendix~\ref{sec:HTL}.

%%%%%%%%%%%%%%%%%%%%%%%%%%%%%%%%%%%%%%%%%%%%%%%%%%%%%%%%%%%%%%%%%%%%%%%%%%%
%
\section{Fundamental Physics of Jet Drift}
\label{sec:slab}
%
%%%%%%%%%%%%%%%%%%%%%%%%%%%%%%%%%%%%%%%%%%%%%%%%%%%%%%%%%%%%%%%%%%%%%%%%%%%

%||||||||||||||||||||||||||||||||||||||||||||||||||||||||||||||||||||||||||
%
\subsection{General Framework and Kinematics}
\label{sec:moments}
%
%||||||||||||||||||||||||||||||||||||||||||||||||||||||||||||||||||||||||||

Our starting point is the expression derived in Ref.~\cite{Sadofyev:2021ohn} for the final-state distribution of jets at first order in opacity, including for the first time the corrections due to the medium motion.   That expression, applied to the case of photon-jet production which is initially back-to-back as in \eq{e:gamjet0}, reads
\begin{align}   \label{e:gen0}
    \frac{dN^{(1)}}{d^3 p_\gamma \, d^3 p} 
    &= \int_0^L \frac{dt}{\lambda(t)} \, \int d^2q_T \: {\hat\sigma}(q^2_T, t)
    \Bigg[\left( 
    \frac{dN^{(0)}}{d^3 p_\gamma \: d^2 (p-q)_T \, dE} \right) \bigg( 1
    + \vec{u}_{T} (t) \cdot  \vec{\Gamma} (\vec{q}_T, t) \bigg)
    \notag\\& \hspace{1cm}
    - \left( \frac{dN^{(0)}}{d^3 p_\gamma \, d^2 p_{T} \, dE} \right) \bigg( 1 + \vec{u}_{T} (t) \cdot  \vec{\Gamma}_{DB} (\vec{q}_T, t) \bigg)\Bigg]\, ,
\end{align}
with $\vec{u}_T$ the component of the fluid velocity $\vec{u}$ transverse to the initial jet momentum and the newly-derived corresponding velocity corrections
\begin{subequations}    \label{e:gen1}
\begin{align}
    \vec{\Gamma} (\vec{q}_T, t) &=
    - 2 \frac{\vec{p}_T - \vec{q}_T}{(1-u_\parallel (t))E}
    + \frac{\vec{q}_T}{(1-u_\parallel (t)) E} \:
    \left( \frac{(p-q)_T^2 - p_T^2}{\hat\sigma(q_T^2, t)} \right) \: 
    \frac{\partial\hat\sigma}{\partial q_T^2}
    \notag \\ & \hspace{1cm}
    - \frac{\vec{q}_T}{1-u_\parallel (t)} \left( \frac{1}{\bar{N}_0 (E, \vec{p}_T - \vec{q}_T)} \frac{\partial \bar{N}_0}{\partial E} \right) \, ,
    \\
    \vec{\Gamma}_{DB}(\vec{q}_T, t) &=
    -2\frac{\vec{p}_T}{(1-u_\parallel (t))E} - \frac{\vec{p}_T}{(1-u_\parallel (t)) E} \: \frac{q_T^2}
    {\hat{\sigma}(q_T^2, t)} \:\frac{\partial\hat{\sigma}}{\partial q_T^2} \, .
\end{align}
\end{subequations}
Eq.~\eqref{e:gen0} is expressed as a line integral over the trajectory of the jet from its jet production point at $t=0$ until it escapes the medium (either at finite distance $L$ or as $L \rightarrow \infty$), with the mean free path $\lambda(t)$, normalized elastic cross section $\hat\sigma(q_T^2 , t) = \frac{1}{\sigma^\mathrm{el}_\mathrm{tot}} \frac{d\sigma^\mathrm{el}}{d^2 q_T}$, and velocity $\vec{u}(t)$ varying along the trajectory.  The exchange of transverse momentum $\vec{q}_T$ with the medium results in the broadening and drift of the jet distribution.  Since the jet is highly boosted along its initial direction of motion, there are distinct effects from vector components parallel to the direction of motion versus perpendicular to it.  For an arbitrary 3-vector $\vec{V}$, the projections may be expressed compactly in terms of the unit vector $\hat{n}_{J, i} = - \vec{p}_\gamma / E_\gamma$ in the initial jet direction (that is, back-to-back with the photon):
\begin{subequations}    \label{e:paraperp}
\begin{align}
    V_\parallel &\equiv
    \vec{V} \cdot \hat{n}_{J, i} \: ,
    \\
    \Tvec{V} &\equiv 
    \vec{V} - V_\parallel \, \hat{n}_{J, i} \: .
\end{align}
\end{subequations}
The velocity corrections $\vec{\Gamma}, \vec{\Gamma}_{DB}$ arise primarily from two physical effects: a shift in the momentum scale entering the cross section $\hat{\sigma}$ and a shift in the jet energy in the initial distribution $\bar{N}_0 \equiv E_\gamma E \frac{dN}{d^3 p_\gamma \, d^3 p}$ relative to the measured final-state jet.  These shifts lead to the logarithmic derivatives of $\hat\sigma$ and $\bar{N}_0$ as expressed in Eqs.~\eqref{e:gen1}.

The projections $\eqref{e:paraperp}$ form a natural coordinate system in which to express the interaction of the jet with the medium, so that for on-shell momenta like $p_\gamma^\mu$ we have $E_\gamma^2 = \vec{p}_\gamma^2 = p_{\gamma \, \parallel}^2 + p_{\gamma \, T}^2$, and similarly for $p^\mu$ (working in the massless limit).  Using these components to express the $\gamma + \mathrm{jet}$ double distribution gives
\begin{subequations}
\begin{align}
    \frac{dN}{d^3 p_\gamma \, d^2 p_\T \, dE}
    &=
    \left| \frac{\partial p_\parallel}{\partial E} \right|
    \: \frac{dN}{d^3 p_\gamma \, d^3 p}
    \approx \frac{dN}{d^3 p_\gamma \, d^3 p} \: ,
    \\
    \frac{dN^{(0)}}{d^3 p_{\gamma} \: d^2 p_\T \, d E} &=
    \frac{dN^{(0)}}{d^3 p_\gamma} \:
    \delta^2 (\vec{p}_\T) 
    \, \delta(E - E_\gamma) \: ,
\end{align}
\end{subequations}
where for $p_\parallel = E \sqrt{1 - \frac{p_\T^2}{E^2}} \approx E \left(1 + \ord{\frac{p_\T^2}{E^2}}\right)$ the Jacobian $\left| \frac{\partial p_\parallel}{\partial E} \right|$ is approximately unity as long as the acquired transverse momentum $p_\T$ is small compared to the jet energy $E$.
Substituting these expressions into the distribution \eq{e:gen0} at first order in opacity, we obtain
\begin{align}
    \frac{dN^{(1)}}{d^3 p_\gamma \: d^2 p_{\T} \, dE} 
    &= 
    \frac{dN^{(0)}}{d^3 p_\gamma} \:
    \delta(E - E_\gamma)
    \int \frac{dt}{\lambda(t)} \, d^2q_\T \: {\hat\sigma}(q^2_T, t)
    \notag \\ & \hspace{-1cm} \times
    \Bigg[\delta^2(\Tvec{p} - \Tvec{q}) 
    \bigg( 1 + \vec{u}_{\T} (t) \cdot  \vec{\Gamma} (\vec{q}_\T, t) \bigg)
    -\delta^2(\Tvec{p})
    \bigg( 1 + \vec{u}_{\T} (t) \cdot  \vec{\Gamma}_{DB} (\vec{q}_\T, t) \bigg)\Bigg]\, .
\end{align}
Since the primary distribution of interest is the distribution of the jet acoplanarity $\Delta\theta \approx p_\T / E$, we integrate the photon momentum $\vec{p}_\gamma$ over some bin defining the typical photon direction and energy:
\begin{subequations}
\begin{align}
    \int\limits_\mathrm{bin} \, d^3 p_\gamma \, \frac{dN^{(0)}}{d^3 p_\gamma} &=
    \int\limits_\mathrm{bin} \, d^3 p_\gamma \, \int d^3 p \, \frac{dN^{(0)}}{d^3 p_\gamma \, d^3 p}
    = \frac{
        \sigma_\mathrm{bin}^{\gamma + \mathrm{jet}} }{
        \sigma_\mathrm{tot}^{\gamma + \mathrm{jet}} } ,
    \\
    \frac{dN_\mathrm{bin}}{d^2 p_\T \, dE} &\equiv
    \frac{1}{   \sigma_\mathrm{bin}^{\gamma + \mathrm{jet}} \, / \,
        \sigma_\mathrm{tot}^{\gamma + \mathrm{jet}} } \:
    \int\limits_\mathrm{bin} \, d^3 p_\gamma \, \int d^3 p \,
    \frac{dN}{d^3 p_\gamma \: d^2 p_\T \, dE} ,
\end{align}
\end{subequations} 
giving
\begin{align}   \label{e:gen2}
    \frac{dN_\mathrm{bin}^{(1)}}{d^2 p_{\T} \, dE} 
    &= 
    \delta(E - E_\gamma)
    \int \frac{dt}{\lambda(t)} \, d^2q_\T \: {\hat\sigma}(q^2_T, t)
    \notag \\ & \hspace{-1cm} \times
    \Bigg[\delta^2(\Tvec{p} - \Tvec{q}) 
    \bigg( 1 + \vec{u}_{\T} (t) \cdot  \vec{\Gamma} (\vec{q}_\T, t) \bigg)
    -\delta^2(\Tvec{p})
    \bigg( 1 + \vec{u}_{\T} (t) \cdot  \vec{\Gamma}_{DB} (\vec{q}_\T, t) \bigg)\Bigg]\, .
\end{align}
In the limit where the bin size becomes infinitely small, this can be interpreted as ``dividing'' by the distributions $\frac{dN^{(0)}}{d^3 p_\gamma}$.  Hereafter we will drop the subscript ``$\mathrm{bin}$'' as understood.  

The second term of \eq{e:gen2} arises from the ``double-Born'' diagrams and reflects the depletion of the original back-to-back peak at $\Delta \theta = \pi$ due to out-scattering into the continuous distribution of $\Tvec{p}$.  Since this term contributes only to final states with $\Tvec{p} = 0$ exactly, it is irrelevant for the continuous part of the distribution, and we will drop it going forward.  We also integrate \eq{e:gen2} over the jet energy $E$, picking up the delta function and leaving just the continuous distribution
\begin{align}   \label{e:gen3}
    \frac{dN^{(1)}}{d^2 p_\T} 
    &= 
    \int \frac{dt}{\lambda(t)} \: {\hat\sigma}(p^2_T, t) \:
    \bigg( 1 + \Tvec{u} (t) \cdot \vec{\Gamma} (\Tvec{p}, t) \bigg) 
    \notag \\ & \hspace{-0.5cm}=
    \int \frac{dt}{\lambda(t)} \: {\hat\sigma}(p^2_T, t) \:
    \bigg[ 1 
    - \frac{\Tvec{u} (t) \cdot \Tvec{p}}{1-u_\parallel (t)} \:
    \bigg(
    \frac{1}{E} \:
    \frac{p_\T^2}{\hat\sigma(p_\T^2, t)} \: 
    \frac{\partial\hat\sigma}{\partial p_\T^2}
    + \frac{1}{\bar{N}_0 (E, \Tvec{0})} \frac{\partial \bar{N}_0}{\partial E}
    \bigg)
    \bigg]
\end{align}
for any $\Tvec{p} \neq 0$.  For definiteness, let us explicitly insert the Gyulassy-Wang potential \cite{Gyulassy:1993hr} and tree-level initial jet distribution arising from the hard scattering
\begin{subequations}    \label{e:GWsub}
\begin{align}
    \hat\sigma(q_\T^2, t) &= \frac{1}{\pi} \: \frac{\mu^2 (t)}{(q_\T^2 + \mu^2(t))^2} & \rightarrow &&
    \frac{1}{\hat{\sigma} (q_\T^2, t)} \frac{\partial \hat\sigma}{\partial q_\T^2} &= 
    \frac{-2}{q_\T^2 + \mu^2(t)} \: ,
    \\
    \bar{N}_0 (E, \Tvec{0}) &\propto \frac{1}{E^4}  &   \rightarrow &&
    \frac{1}{\bar{N}_0 (E, \Tvec{0})} \frac{\partial \bar{N}_0}{\partial E} &= \frac{-4}{E} \:,
\end{align}
\end{subequations}
with $\mu$ the Debye mass, giving
\begin{align}   \label{e:gen4}
    \frac{dN^{(1)}}{d^2 p_\T} 
    &= 
    \frac{1}{\pi} \int \frac{dt}{\lambda(t)} \: \frac{\mu^2 (t)}{(p_\T^2 + \mu^2(t))^2} \:
    \left[ 1 + \frac{\Tvec{u} (t) \cdot \Tvec{p}}{\left( 1-u_\parallel (t) \right) E} \:
    \bigg( \frac{6 p_\T^2 + 4 \mu^2 (t) }{p_\T^2 + \mu^2 (t)} \bigg) \right].
\end{align}
While for most of this paper we work with the Gyulassy-Wang potential as in \eqref{e:GWsub}, we explore the sensitivity of our effect to this choice in Appendix~\ref{sec:HTL} by explicit comparison to the Hard Thermal Loop framework. We note that the first term, $1$, in brackets in \eq{e:gen4} is the usual symmetric jet broadening distribution at first order in opacity, while the second, asymmetric term represents the jet drift effect which skews the distribution in the direction of the velocity $\vec{u}$. 

\begin{figure}
\begin{centering}
\includegraphics[width=0.48\textwidth]{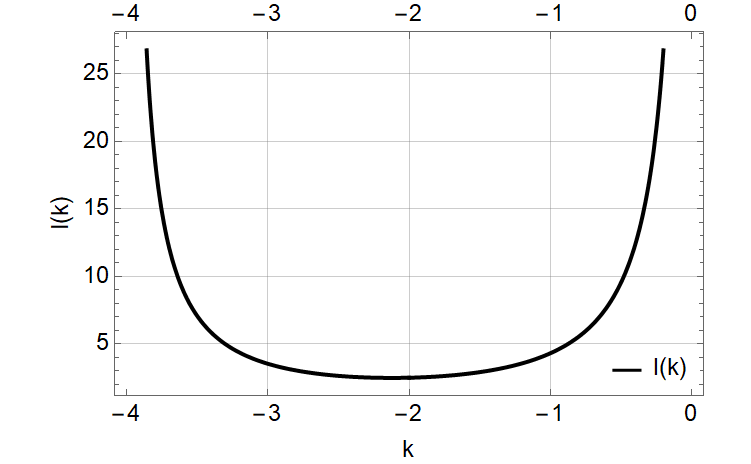}
\includegraphics[width=0.48\textwidth]{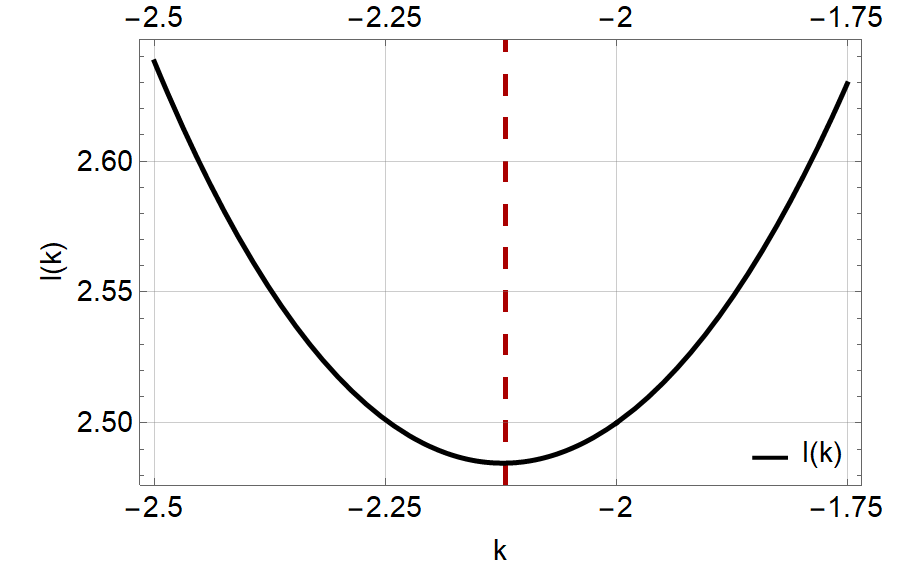}
\caption{Left: Plot of the coefficient $I(k)$. Right: Zoom displaying the minimum of $I(k)$ to demonstrate its skewness.}
\label{f:prefactor}
\end{centering}
\end{figure}

As discussed in Sec.~\ref{sec:intro}, the usual symmetric broadening of the acoplanarity distribution is difficult to discern against a background of many competing symmetric mechanisms, while the antisymmetric jet drift effect has few counterparts.  We single out the jet drift term of \eq{e:gen4} by computing odd (vector-valued) moments of the distribution:
\begin{align}   \label{e:vecmoment1}
    \left\langle \Tvec{p} \, p_\T^k \right\rangle \equiv
    \frac{
        \int d^2 p_\T \, (\Tvec{p} \, p_\T^k) \, \frac{dN}{d^2 p_\T}
    }{
        \int d^2 p_\T \, \frac{dN}{d^2 p_\T}
    }
    \approx     \int d^2 p_\T \, (\Tvec{p} \, p_\T^k) \, \frac{dN^{(1)}}{d^2 p_\T} \:,
\end{align}
where $k$ is an arbitrary power, and in the last step we have evaluated the numerator and denominator to the first nonvanishing order in opacity (first and zeroth order, respectively).  The explicitly vector-valued nature of the antisymmetric moment \eqref{e:vecmoment1} makes clear why the usual static result -- along with any other symmetric backgrounds -- cannot contribute: they integrate to zero.  Only the presence of an external vector direction (here provided by the velocity $\vec{u}$) can produce a nonzero vector-valued moment $\eqref{e:vecmoment1}$:
\begin{align}   \label{e:vecmoment2}
    \left\langle \Tvec{p} \, p_\T^k \right\rangle &= 
    \frac{1}{\pi \, E} \int \frac{dt}{\lambda(t)} \: 
    \frac{\vec{u}_{\T i} (t)}{1-u_\parallel (t)} \:
    \mu^2 (t) \: \left[
    \int d^2 p_\T \, (\Tvec{p} \, p_\T^k) \:
    \vec{p}_{\T i} \:
    \frac{6 p_\T^2 + 4 \mu^2 (t) }{(p_\T^2 + \mu^2 (t))^3}  \right]
    \notag \\ &=
    \frac{1}{E} \int \frac{dt}{\lambda(t)} \: 
    \frac{\Tvec{u} (t)}{1-u_\parallel (t)} \:
    \mu^2 (t) \: \left[
    \int d p_\T^2 \, (p_\T^2)^{1+k/2} \:
    \frac{3 p_\T^2 + 2 \mu^2 (t) }{(p_\T^2 + \mu^2 (t))^3}  \right]
    \notag \\ &=
    \frac{I(k)}{E} \int \frac{dt}{\lambda(t)} \: 
    \frac{\Tvec{u} (t)}{1-u_\parallel (t)} \:
    \mu^{k+2} (t)
\end{align}
with the $k$-dependent prefactor
\begin{align}   \label{e:prefactor}
    I(k) \equiv \int_0^\infty d \xi \, \xi^{1+k/2} \:
    \frac{3 \xi + 2}{(\xi + 1)^3} =
    - \frac{\pi}{8} (k+2)(k+12) \, \csc(\tfrac{k}{2}\pi)
\end{align}
plotted in Fig.~\ref{f:prefactor}.

%--------------------------------------------------------------------------
%--------------------------------------------------------------------------

\begin{figure}
\begin{centering}
\includegraphics[height=0.49\textwidth]{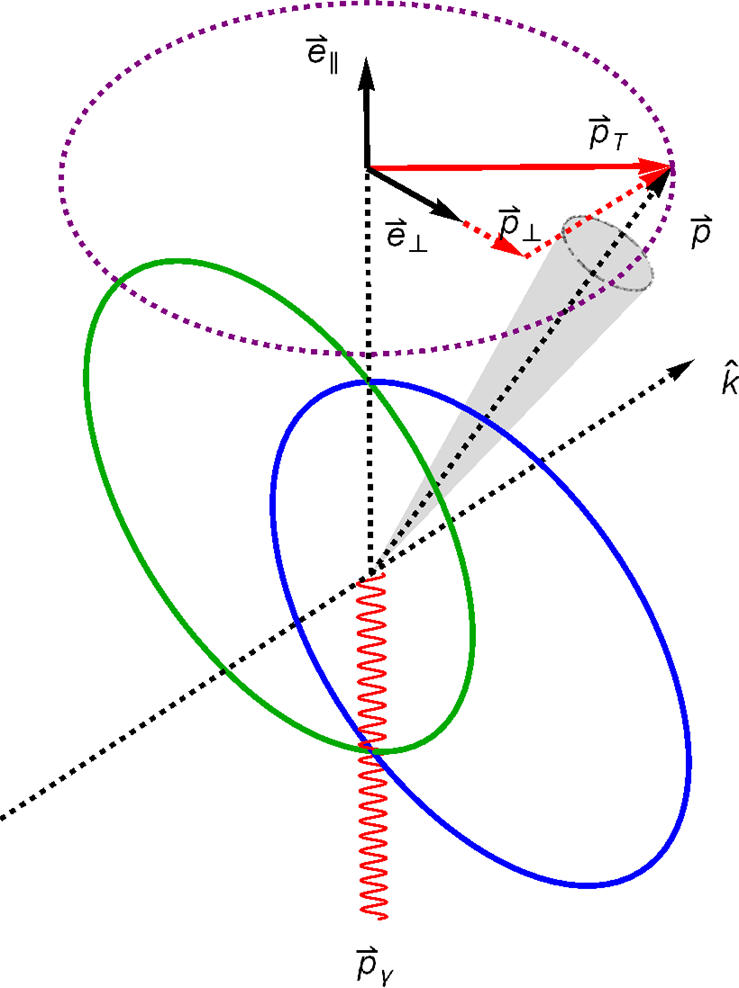} \hspace{0.04cm}
\includegraphics[width=0.45\textwidth]{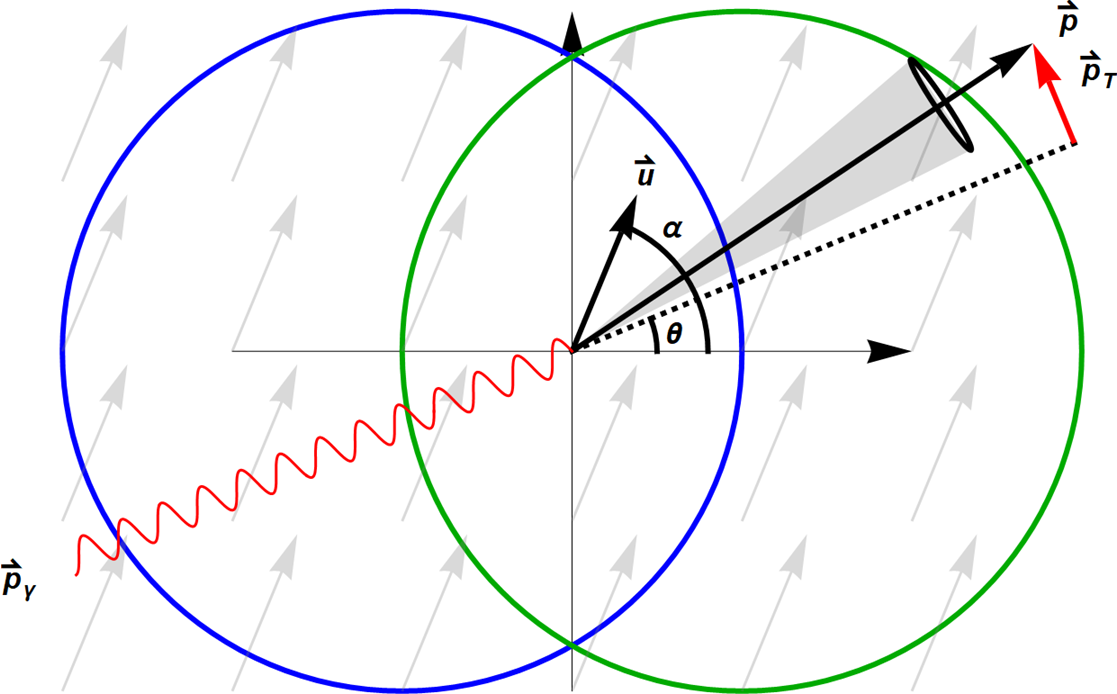}
\caption{Left:  Graphic depiction of jet coordinate system, with beam axis along $\hat{k}$. 
Right:  Two-dimensional coordinates in the $xy$ plane.
}
\label{f:coords}
\end{centering}
\end{figure}

For the purposes of this paper, let us restrict ourselves to the case of $\gamma + \mathrm{jet}$ production with the photon being produced at mid rapidity, such that the initial jet direction $\hat{n}_{J, i}$ lies in the $xy$ plane transverse to the beam axis ($z$).  Then the transverse vectors $\vec{p}_T, \vec{u}_T$ associated with the $\gamma + \mathrm{jet}$ acoplanarity generally lie in the plane orthogonal to $\hat{n}_{J, i}$ and may include a $z$ component along the beam axis.  We will further assume that the velocity profile at mid rapidity is boost invariant, so that $\vec{u}$ has no $z$ component.  Then, even though the symmetric broadening effect of the jet direction includes a broadening of its rapidity away from the photon rapidity, the preferred direction which skews the acoplanarity distribution lies entirely within the $xy$-plane.  We can make these kinematics explicit by introducing basis vectors in cylindrical coordinates,

\begin{subequations}
\begin{align}
    \hat{e}_\parallel &=\cos\theta \, \hat{i} + \sin\theta \, \hat{j} = \hat{n}_{J, i} \, , \\
    \hat{e}_\PrpXY &=-\sin\theta \, \hat{i} + \cos\theta \, \hat{j} = \hat{k} \times \hat{n}_{J, i} \, , \\
    \vec{u} &= u_\parallel \, \hat{e}_\parallel + u_\PrpXY \, \hat{e}_\PrpXY \, , \\
    \vec{p} &= p_\parallel \, \hat{e}_\parallel + p_\PrpXY \, \hat{p}_\PrpXY + p_z \, \hat{k} \, , \\
    \tvec{u} \cdot \tvec{p} &= u_\PrpXY p_\PrpXY \, ,
\end{align}
\end{subequations}
as visualized in Fig. \ref{f:coords}. Here, the initial jet direction $\hat{e}_\parallel$ makes an azimuthal angle of $\theta$ with respect to the $+x$-axis.
%
%--------------------------------------------------------------------------
%--------------------------------------------------------------------------
%
This makes the one-dimensional nature of the kinematics explicit; with the $\gamma + \: \mathrm{jet}$ pair being produced at mid-rapidity and a boost-invariant assumption for the flow field $\vec{u}_T = u_\bot \hat{e}_\bot$, the jet drift lies entirely in the azimuthal direction $\hat{e}_\bot$:
\begin{align}   \label{e: moment}
    \left\langle \tvec{p} \, p_\PrpXY^k \right\rangle &= 
    \frac{I(k)}{E} \, \hat{e}_\PrpXY \, \int \frac{dt}{\lambda(t)} \: 
    \frac{u_\PrpXY (t)}{1-u_\parallel (t)} \:
    \mu^{k+2} (t) \, .
\end{align}

We can now employ \eq{e: moment} directly to compute various odd moments of the acoplanarity distribution in a range of geometries through the line integral \eqref{e: moment}.  The moments $\left\langle \tvec{p} \, p_\PrpXY^k \right\rangle$ describe the asymmetric shift in the jet distribution with different weight on the UV and IR momentum regions depending on the choice of index $k$.  We can study any of these moments for values of $k$ which are convergent; as we will see in Appendix ~\ref{sec:HTL}, the range of convergence of these moments depends on the choice of potential.  As such, exploring various experimental measures of asymmetries like $\left\langle \tvec{p} \, p_\PrpXY^k \right\rangle$ in the acoplanarity distribution can help discriminate between different models of the medium potential, which must all agree in the UV regime but may differ in the IR. Of particular interest is the moment $k=0$ which corresponds simply to the mean acoplanarity \eqref{e:meanvec1}:
\begin{align}  
    \langle \Delta\theta \rangle \, \hat{e}_\PrpXY = \frac{\langle \tvec{p} \rangle}{E} =    
    I(0) \, \hat{e}_\PrpXY \,
    \int\frac{dt}{\lambda(t)} \, \frac{u_\PrpXY(t)}{1 - u_\parallel (t)} \, \frac{\mu^2 (t)}{E^2} .
\end{align}
The prefactor $I(k)$ from \eqref{e:prefactor} is logarithmically divergent in the UV for $k = 0$, so we must regulate the upper limit of $p_\PrpXY$ by a cutoff of order $p_{\bot \, , \, max} \sim\ord{\sqrt{E \mu}}$ rather than extending the integral to infinity \cite{Gyulassy:2000er}.  Doing so gives
\begin{align}
    I(0) &= \!\!\! \int\limits_1^{E / \mu(t)} \!\!\!
    d \xi \, \xi \:
    \frac{3 \xi + 2}{(\xi + 1)^3}
    \approx
    3 \ln\frac{E}{\mu(t)}
\end{align}
within the leading-logarithmic approximation, such that
\begin{align}   \label{e:meantheta1}
    \langle \Delta\theta \rangle \, = 3 \int\frac{dt}{\lambda(t)} \, \frac{u_\PrpXY(t)}{1 - u_\parallel (t)} \, \frac{\mu^2 (t)}{E^2} \, \ln\frac{E}{\mu(t)} .
\end{align}
A more detailed analysis of the moments \eqref{e: moment} and \eqref{e:meantheta1} in increasingly sophisticated geometries is the main subject of the rest of this paper.

%||||||||||||||||||||||||||||||||||||||||||||||||||||||||||||||||||||||||||
%
\subsection{Jet Drift in the Constant Slab Approximation}
\label{sec:slabmoments}
%
%||||||||||||||||||||||||||||||||||||||||||||||||||||||||||||||||||||||||||

As a starting point to identify the fundamental physics of jet drift, let us begin by evaluating the moments \eqref{e: moment} for the case of a constant slab of plasma of length $L$:
\begin{align} \label{e: simplemoment}
    \langle \vec{p}_\PrpXY p_\PrpXY^k \rangle=\frac{\vec{u}_\PrpXY}{1-u_\parallel}\frac{L}{\lambda}\frac{\mu^{k+2}}{E}I(k) \, ,
\end{align}
or equivalently,
\begin{align}
    \langle \vec{p}_\PrpXY p_\PrpXY^k \rangle=\frac{(\vec{u}\cdot\hat{e}_\PrpXY)\hat{e}_\PrpXY}{1-(\vec{u}\cdot\hat{e}_\parallel)}\frac{L}{\lambda}\frac{\mu^{k+2}}{E}I(k) .
\end{align}
Taking the fluid velocity $\vec{u}$ to make an azimuthal angle $\alpha$ with respect to the $+x$-axis, as depicted in the right panel of Fig. \ref{f:coords}, an expression for the magnitude of the moment in terms of fluid velocity (magnitude $u$ and direction $\alpha$) and jet angle $\theta$ can be constructed:
\begin{align}  \label{e: fluidvelocityangle}
     \left|\langle p_\PrpXY p_\PrpXY^k \rangle\right|=\frac{-u\sin{(\theta-\alpha)}}{1-u\cos{(\theta-\alpha)}}\frac{L}{\lambda}\frac{\mu^{k+2}}{E}I(k). 
\end{align} 
%

%--------------------------------------------------------------------------
\begin{figure}
\begin{centering}
\includegraphics[width=0.49\textwidth]{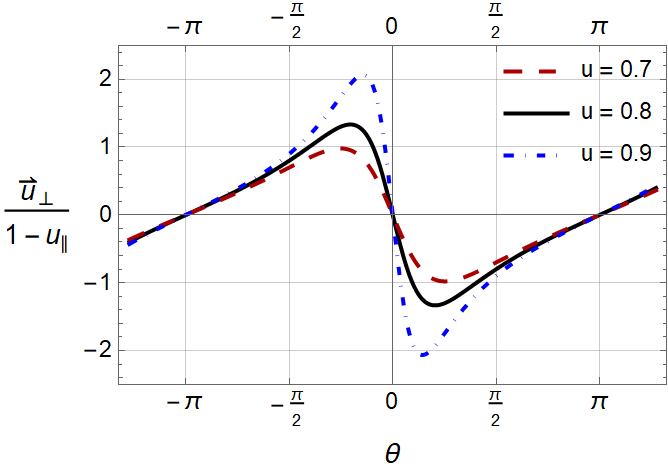}
\includegraphics[width=0.49\textwidth]{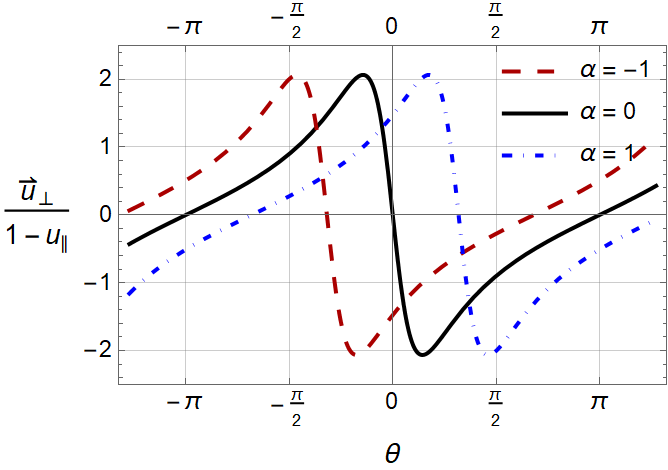}
\caption{Plot of moment geometry dependence term. Left panel: Plot for $\alpha=0$ illustrating the dependence on the fluid speed $u$.  Right panel: Plot for $u = 0.9$ illustrating the dependence on the fluid direction $\alpha$. Constant factors have been divided out.
\label{f:MA1}
}
\end{centering}
\end{figure}
%--------------------------------------------------------------------------

\begin{figure}
\begin{centering}
\includegraphics[width=0.49\textwidth]{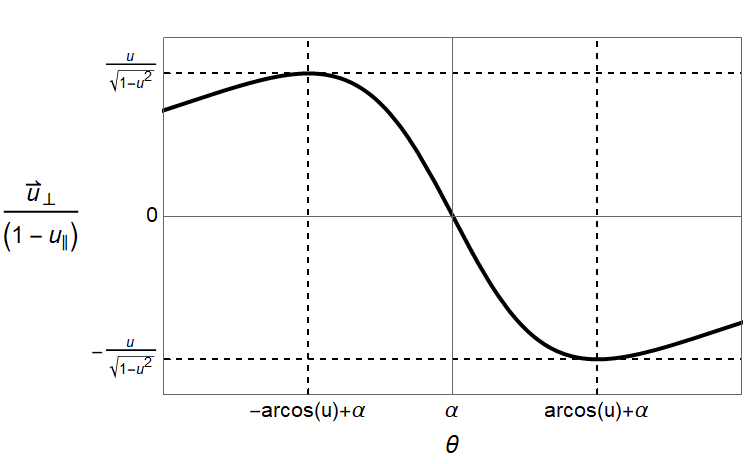}
\includegraphics[width=0.49\textwidth]{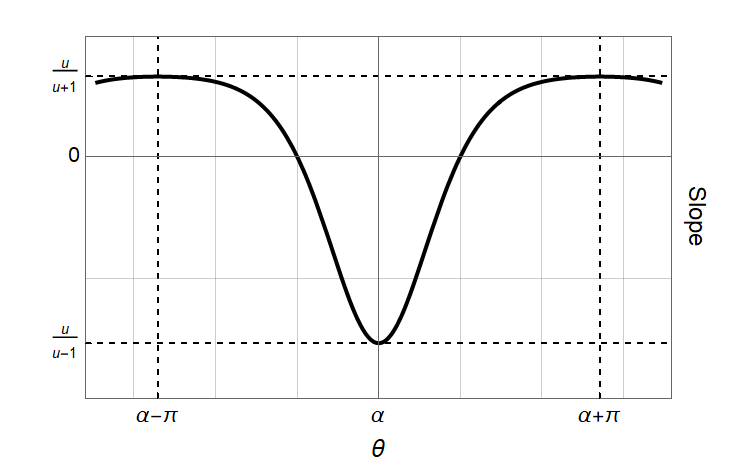}
\caption{Graphic depiction of tomographic velocity information encoded in the structure of the geometric coupling.
\label{f:tomographicInfo}
}
\end{centering}
\end{figure}

The geometry dependence of Eq.~\eqref{e: fluidvelocityangle} is plotted in Fig.~\ref{f:MA1} for various values of the fluid direction $\alpha$ and fluid speed $u$.  The geometry dependence of the moments is contained in the factor $u_\PrpXY / (1 - u_\parallel)$, leading to the characteristic pattern of zero crossings and extrema seen in Fig.~\ref{f:tomographicInfo} which encodes the tomographic information about the medium flow.
%--------------------------------------------------------------------------
%--------------------------------------------------------------------------
The zero crossings' locations at $\theta=\alpha$ and $\theta=\alpha\pm\pi$ can be explained by the fact that at those angles, the jet is travelling either parallel to or antiparallel to the flow velocity, leading to $u_\PrpXY=0$ and thus $\langle p_\PrpXY p_\PrpXY^k \rangle=0$.  The two extrema, which increase in magnitude for larger fluid speed $u$, are located by finding the critical points of the moment with respect to $\theta$, leading to
\begin{align}   \label{e:thetapeak}
    \theta_\mathrm{peak}=\alpha\pm\arccos{u} \, .
\end{align}
The tomographic information about the flow velocity is also encoded in the moments moments \eqref{e: simplemoment} in other ways. For instance, substituting \eq{e:thetapeak} into \eq{e: fluidvelocityangle} gives the dependence of extrema height (whether maximum or minimum) on the fluid speed,
\begin{align} \label{e: peak}
    \left|\langle p_\PrpXY p_\PrpXY^k \rangle\right|_\mathrm{peak} = \frac{u}{\sqrt{1-u^2}}\frac{L}{\lambda}\frac{\mu^{k+2}}{E}I(k) \, ,
\end{align}
which increases with increasing speed $u$ and diverges as $u \rightarrow 1$.  The fluid speed can also be extracted from the slopes of the curve at the zero crossings:
\begin{subequations}    \label{e:slopes}
\begin{align} \label{e:slope1}
    m(\theta=\alpha) &\equiv \frac{d}{d\theta} \langle p_\PrpXY p_\PrpXY^k \rangle \bigg|_{\theta=\alpha}=\frac{u}{u-1}\frac{L}{\lambda}\frac{\mu^{k+2}}{E}I(k) \, ,
    \\  \label{e:slope2}
    m(\theta=\alpha\pm\pi) &\equiv \frac{d}{d\theta} \langle p_\PrpXY p_\PrpXY^k \rangle \bigg|_{\theta=\alpha\pm\pi}=\frac{u}{u+1}\frac{L}{\lambda}\frac{\mu^{k+2}}{E}I(k) \, .
\end{align}
\end{subequations}
One interesting feature of the slopes is the fact that as $u\to1$, the slope at $\theta = \alpha$ approaches negative infinity, whereas the slope at $\theta = \alpha \pm \pi$ approaches a finite value.  The increase in the slope at $\theta = \alpha$ with increasing $u$ and the constancy of the slope at $\theta = \alpha \pm \pi$ are clearly seen in Fig. \ref{f:MA1}.  Moreover, the ratio of these two slopes cancels out all other dependence on the medium properties (and on the index $k$),  leaving just a dependence on fluid speed:
\begin{align} \label{e:sloperatio}
    \frac{m(\theta=\alpha\pm\pi)}{m(\theta=\alpha)}=\frac{u-1}{u+1} \, .
\end{align}

As we have shown, the position and width of the peak-zero-peak pattern is linked directly to the fluid velocity. The zero crossings show the direction, while the positions of the extrema, the slopes at each zero crossing, and the heights of the peaks encode the magnitude. Even in this simplest possible realization of a jet interacting with a flowing medium, the pattern of jet drift as expressed in the moments \eqref{e: moment} carries a tremendous amount of information about the fluid flow $\vec{u}$.  Each of the curves shown in Fig.~\ref{f:MA1} could be used to completely reconstruct both the magnitude and direction of $\vec{u}$ in multiple ways, suggesting significant promise to the idea of ``jet-flow tomography.''  This significant structure even in the constant slab case gives us reason to think that the moments will continue to carry information about the medium flow even in more complex geometries and flow patterns.

%%%%%%%%%%%%%%%%%%%%%%%%%%%%%%%%%%%%%%%%%%%%%%%%%%%%%%%%%%%%%%%%%%%%%%%%%%%
%
\section{Coupling to Elliptic Flow in Heavy-Ion Collisions}
\label{sec:geometry}
%
%%%%%%%%%%%%%%%%%%%%%%%%%%%%%%%%%%%%%%%%%%%%%%%%%%%%%%%%%%%%%%%%%%%%%%%%%%%

%||||||||||||||||||||||||||||||||||||||||||||||||||||||||||||||||||||||||||
%
\subsection{Elliptical Gaussian Toy Model}
\label{sec:model}
%
%||||||||||||||||||||||||||||||||||||||||||||||||||||||||||||||||||||||||||

\begin{figure}
\begin{centering}
\includegraphics[width=0.48\textwidth]{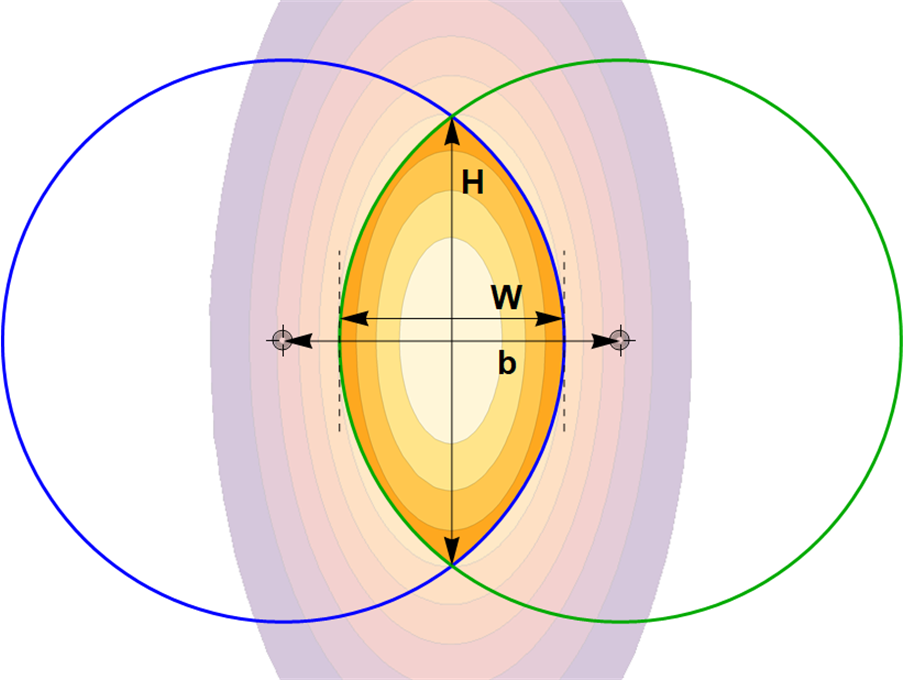}
\includegraphics[width=0.48\textwidth]{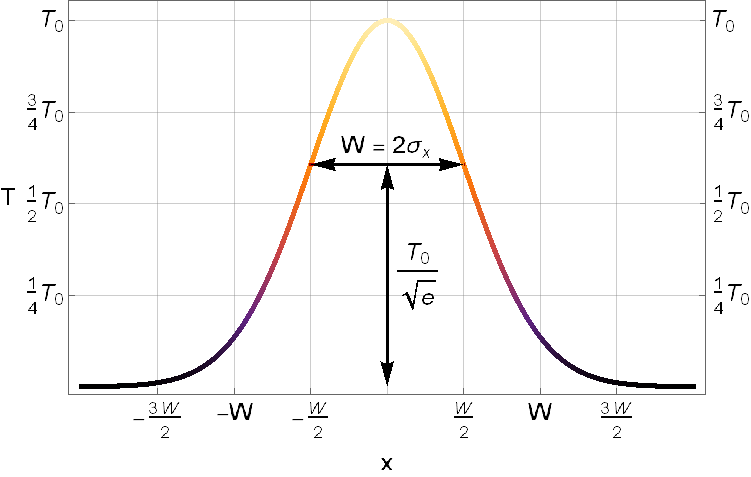}
\caption{Left: Impact parameter relationship with event geometry. Note profile geometry spills over impact region. Right: 2D x-profile view of temperature Gaussian, emphasizing definition of width.}
\label{f:HWB}

\end{centering}
\end{figure}

To study the physics of jet drift under the conditions produced in heavy-ion collisions, we consider a simplified toy model of the elliptical geometry and associated collective flow produced in noncentral events.  We will represent the elliptical temperature profile produced by two colliding nuclei of radius $R$ by a smooth Gaussian distribution
\begin{align}   \label{e:ellipseT}
    T(x,y) = T_0 \, \exp\left[-\frac{x^2}{2 \sigma_x^2}\right] \, \exp\left[-\frac{y^2}{2 \sigma_y^2}\right]
\end{align}
with normalization $T_0$ and with variable height $H = 2 \sigma_x$ and width $W = 2 \sigma_y$ determined from the impact parameter $b$:
\begin{subequations}    \label{e:HW1}
\begin{align}
    W(b) &= 2 \sigma_x = 2 R - b \: , \\
    H(b) &= 2 \sigma_y = \sqrt{4 R^2 - b^2} \: .
\end{align}
\end{subequations}
The simple expressions \eqref{e:HW1} for the height and width of the Gaussian ellipse \eqref{e:ellipseT} are obtained from matching to the height and width of the almond-shaped overlap region of two circular nuclei with center-to-center displacement $b$ along the $x$-axis, as illustrated in Fig.~\ref{f:HWB}.

%-----------------------------------------------------------------------------
%-----------------------------------------------------------------------------

For a simple system in thermal equilibrium (with, for example, all chemical potentials set to zero), the temperature profile \eqref{e:ellipseT} provides the only scale $T(x,y)$, which determines on the basis of dimensional analysis all other relevant thermodynamic quantities at that point.  In particular, for the Debye mass $\mu$ and mean free path $\lambda$ we have
\begin{subequations}
\begin{align}
    \mu &= g T \: , \\
    \lambda &= \frac{\kappa}{T} \: ,
\end{align}
\end{subequations}
which can be used to evaluate the distribution \eqref{e:gen4} and drift moments \eqref{e:vecmoment2} in the thermal background described by Eqs.~\eqref{e:HW1} through the appropriate line integral over the straight-line trajectory $\left( x(t) \, , \, y(t) \right) = \left( x_0 \, , \, y_0 \right) + \left( \cos\theta \, , \, \sin\theta \right) t$.  In addition to the temperature profile \eqref{e:ellipseT}, we also need to specify the associated flow profile $\vec{u}(x,y)$.  Within the scope of this simplified toy model, we assume that the velocity points in the direction of the pressure gradients; that is, in the direction $\vec{u} \propto - \vec\nabla T$.  Under this assumption, we write
\begin{align}   \label{e:ellipseU}
    \vec{u} (x,y) = 
    u_0 \sqrt{\sigma_{y} \sigma_{x}} \, \left( \frac{x}{\sigma_{x}^2} \hat{i} + \frac{y}{\sigma_{y}^2} \hat{j} \right) 
    \exp\left[-\frac{x^2}{2 \sigma_{x}^2}\right] \, \exp\left[-\frac{y^2}{2 \sigma_{y}^2}\right] 
\end{align}
with the dimensionful factor $\sqrt{\sigma_{y} \sigma_{x}}$ chosen to compensate the dimension of the gradient.  With these assumptions, the acoplanarity distribution and drift moments are given by
\begin{subequations}
\begin{align}   \label{e:elpdistr1}
    \frac{dN^{(1)}}{d^2 p_\PrpXY} 
    &= 
    \frac{1}{\pi} \, \frac{g^2}{\kappa} \int dt \, \frac{T^3 (t)}{(p_\PrpXY^2 + g^2 T^2(t))^2} \:
    \left[ 1 + 
    \frac{\vec{p} \cdot \hat{e}_\PrpXY}{E} \:
    \frac{\vec{u} (t) \cdot \hat{e}_\PrpXY}{1 - \vec{u} (t) \cdot \hat{e}_\parallel} \:
    \frac{6 p_\PrpXY^2 + 4 g^2 T^2 (t) }{p_\PrpXY^2 + g^2 T^2 (t)} \right] \: ,
    \\      \label{e:elpmom1}
    \left\langle \tvec{p} \, p_\PrpXY^k \right\rangle &= 
    \frac{I(k)}{E} \, \frac{g^{k+2}}{\kappa} \int dt \, 
    \frac{\vec{u} (t) \cdot \hat{e}_\PrpXY}{1 - \vec{u} (t) \cdot \hat{e}_\parallel} \:
    T^{k+3} (t) \: .
\end{align}
\end{subequations}

The toy model expressed by the elliptical Gaussian profile \eqref{e:ellipseT} and associated velocity profile \eqref{e:ellipseU} is admittedly simplistic, and there are many features of a realistic heavy-ion collision which it does not capture.  First, the use of a smooth profile corresponds to an ``optical Glauber'' model -- a mean field approximation for the nuclear density which neglects the role of event-by-event fluctuations of the nucleon and sub-nucleonic structure.  One important consequence of this optical Glauber model is that the temperature and velocity profiles possess exact mirror symmetry under reflection about the $x$- and $y$-axes which results in the triangularity (and other odd Fourier harmonics) of the profile automatically vanishing.  This exact mirror symmetry is known to be broken by event-by-event fluctuations, which can be considered corrections to the mean-field approximation.  Second, the expedient choice to use a Gaussian ellipse rather than a more realistic overlap of Woods-Saxon nuclear density profiles results in a distribution which falls too smoothly from its central value to zero.  In reality, the temperatures produced from a Woods-Saxon distribution would be nearly constant in the center of the distribution before falling abruptly toward zero at the edges, rather than smoothly decaying to zero.  One consequence of this is that in the more realistic case, the temperature gradients (and therefore the velocities obtained from $\vec{u} \propto - \vec\nabla T$) will be more concentrated at the edges than in the toy model we employ here.  Finally, the use of the profiles \eqref{e:ellipseT} and \eqref{e:ellipseU} as static distributions ignores the important effects of expansion and cooling of the plasma over time.

Nevertheless, there are a number of important systematics of heavy-ion collisions which \textit{are} captured by the simple toy model encoded in \eq{e:ellipseT} and \eq{e:ellipseU}.  The elliptical shape of the mean field geometry captured by \eq{e:ellipseT} is by far the strongest mode of collective flow in the plasma, with higher harmonics being significantly suppressed in comparison. 
In particular, it is true the odd harmonics such as the triangularity are not exactly zero; however, they are still much smaller than the elliptic flow  driven by the mean-field geometry associated with \eq{e:ellipseT} \cite{Luzum:2013yya}.  The elliptical mean-field geometry, determined as a function of impact parameter as in Eqs.~\eqref{e:HW1}, also allows for meaningful centrality binning which can mimic an experimental analysis.  By sampling a random impact parameter and generating the corresponding elliptical geometry, we can systematically study the effect of jets propagating through geometries of different sizes and shapes on an event-by-event basis and analyze the dependence on the centrality class, as in experimental analyses.  Within the context of this simple geometric profile, we can also consistently implement event-by-event fluctuations of the jet production point, appropriately weighted by the binary collision density, and the deflection of the jet itself.  We pursue this event-by-event analysis in Sec.~\ref{sec:fluct} below, beginning for now with an analysis of the properties of the jet drift in a single fixed elliptical geometry as in Eqs.~\eqref{e:ellipseT} and \eqref{e:ellipseU}.

%||||||||||||||||||||||||||||||||||||||||||||||||||||||||||||||||||||||||||
%
\subsection{Geometry Coupling}
\label{sec:staticellipse}
%
%||||||||||||||||||||||||||||||||||||||||||||||||||||||||||||||||||||||||||

%_____________________________________________________________
%
\begin{figure}[!ht]
\begin{centering}
{\includegraphics[width=0.48\textwidth]{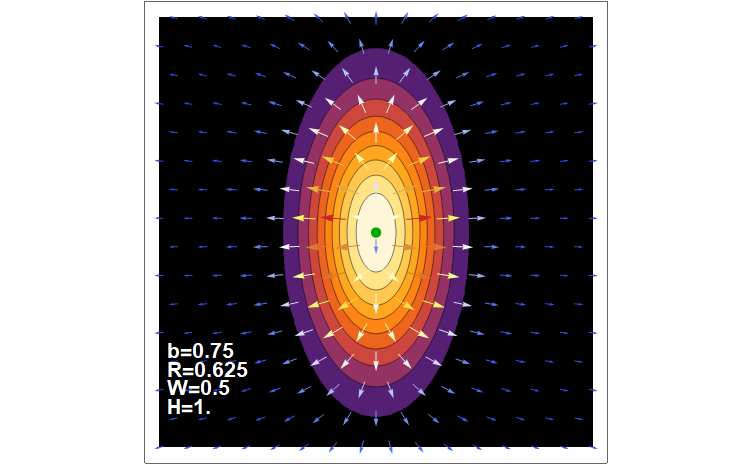}}
\includegraphics[width=0.48\textwidth]{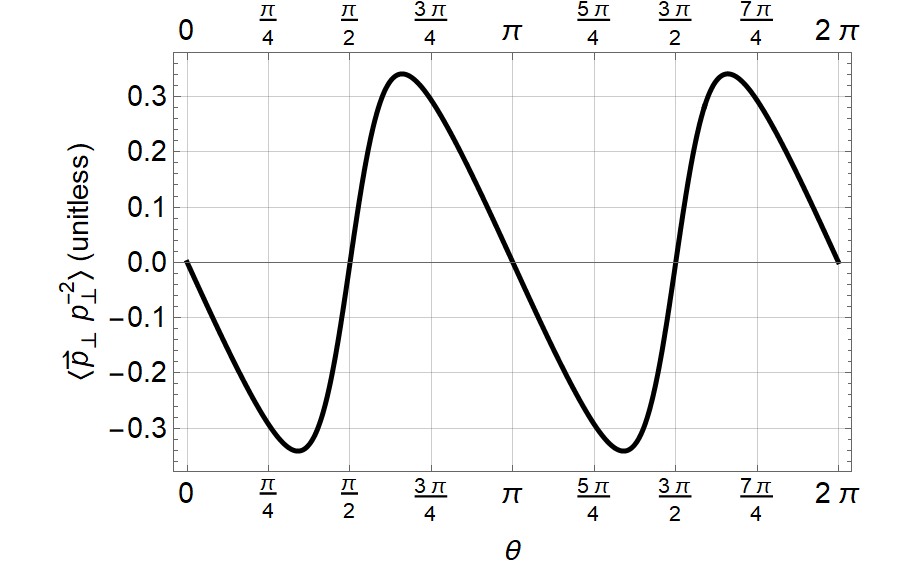}
\caption{Simulated event geometry and corresponding calculated angular profile of moments for production point (green dot) shown at the origin.}
\label{f:EllipseSlice1}
\end{centering}
\end{figure}
%
%_____________________________________________________________

%_____________________________________________________________
%
\begin{figure}[!ht]
\begin{centering}
\includegraphics[width=0.49\textwidth]{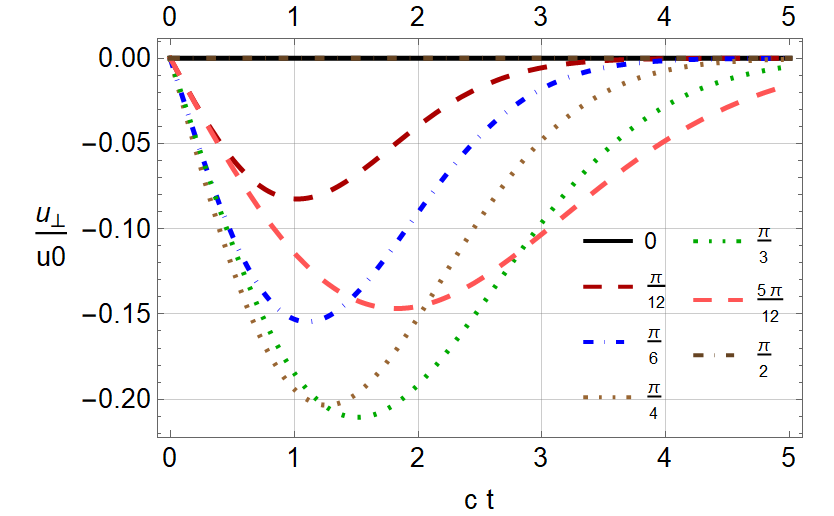}
\includegraphics[width=0.49\textwidth]{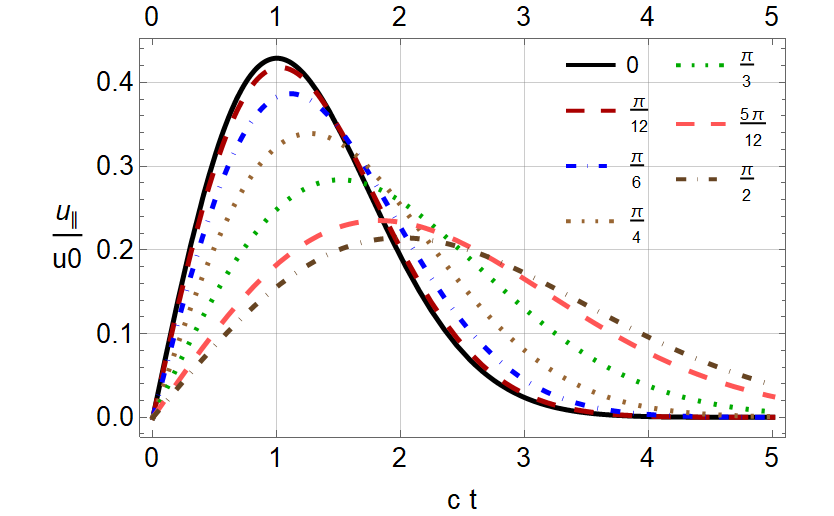}
\caption{Simulated elliptical geometry velocity profiles at jet position as a function of time for several production angles, with jet initially produced at the origin.  Here the geometry parameters are taken to be $W=0.5, H=1$.}
\label{f:EllipseVelocities}

\end{centering}
\end{figure}
%
%_____________________________________________________________

To get a first feel for the response of jet drift to the elliptical geometry described by Eqs.~\eqref{e:ellipseT} and \eqref{e:ellipseU}, consider the drift of a jet produced at the center: $\left( x_0 \, , \, y_0 \right) = (0, 0)$.  We show in Fig.~\ref{f:EllipseSlice1} the particular jet drift moment $k=-2$ obtained by numerical evaluation of the line integral in \eq{e:elpmom1} for various initial jet angles $\theta$, and we further show in Fig.~\ref{f:EllipseVelocities} the velocity components $u_\perp \, , \, u_\parallel$ contributing to the integrand.  The contour plot of the temperature clearly reflects the different path length seen by jets produced along the minor axis of the ellipse $(\theta = 0 \, , \, \pi)$ versus the major axis $(\theta = \pi / 2 \, , \, 3\pi/2)$.  The different axes of the ellipse are also characterized by different gradients, and therefore different magnitudes of the velocity.  As clearly seen in the $u_\parallel$ distribution in Fig. \ref{f:EllipseVelocities}, the radial flow along the minor axis of the ellipse is larger and occurs at earlier times in the jet trajectory compared with the radial flow along the major axis, which is smaller in magnitude and occurs later on.  

By far the most important contribution to the overall moment is the transverse velocity component $u_\PrpXY$.  One immediate and striking feature is the sign pattern of $u_\PrpXY (\theta)$, which is negative definite in the first quadrant $0 < \theta < \pi / 2$, positive in the second quadrant $\pi / 2 < \theta < \pi$, and continuing to alternate.  This feature is solely a function of the elliptical geometry itself, reflecting the fact that the gradients (and hence the flow) are mostly horizontal, tending to drag the jet toward the $\pm x$ axis from either direction.  The magnitude of the transverse flow $u_\PrpXY$ is identically zero along any of the major/minor axes, since there the flow is exactly parallel to the jet, and it rises to a maximum at around $\theta \approx \pm \pi / 3, \pm 2\pi /3$ before decreasing back to zero again, as we see in Fig.~\ref{f:EllipseVelocities}.

This geometry-driven pattern of alternating positive and negative signs for the jet drift moment is another reflection of the attractor physics of jet drift discussed in Appendix~\ref{sec:trajectory}.  The fluid flow, directed mostly along the directions $\theta = 0, \pi$ of the minor axis, acts as an attractor for the jet trajectories, with the sign pattern in each quadrant leading to a systematic deflection of the jet toward $\theta = 0, \pi$ from both above and below.  Correspondingly, the major axes of the ellipse $\theta = \pi/2 , 3\pi / 2$ behave as repulsors of the trajectories, pushing the jet either to the left or the right into the $\theta = 0, \pi$ basins of attraction.  This coupling of the jet drift moments to the elliptical shape of the geometry can be seen in more than just the sign pattern in each quadrant; the full angular dependence of the moments $\langle \tvec{p} p_\PrpXY^k \rangle$ in Fig.~\ref{f:EllipseSlice1} evinces an extremely strong $\cos2\theta$ modulation which dominates the angular dependence.  This plot reveals a central feature which will persist throughout all of our subsequent analyses of jet drift in a fluctuating medium: the strong coupling of the acoplanarity moments to the medium geometry through the orientation of the reaction plane.

%_____________________________________________________________
%
\begin{figure}[!ht]
\begin{centering}
\includegraphics[width=0.5\textwidth]{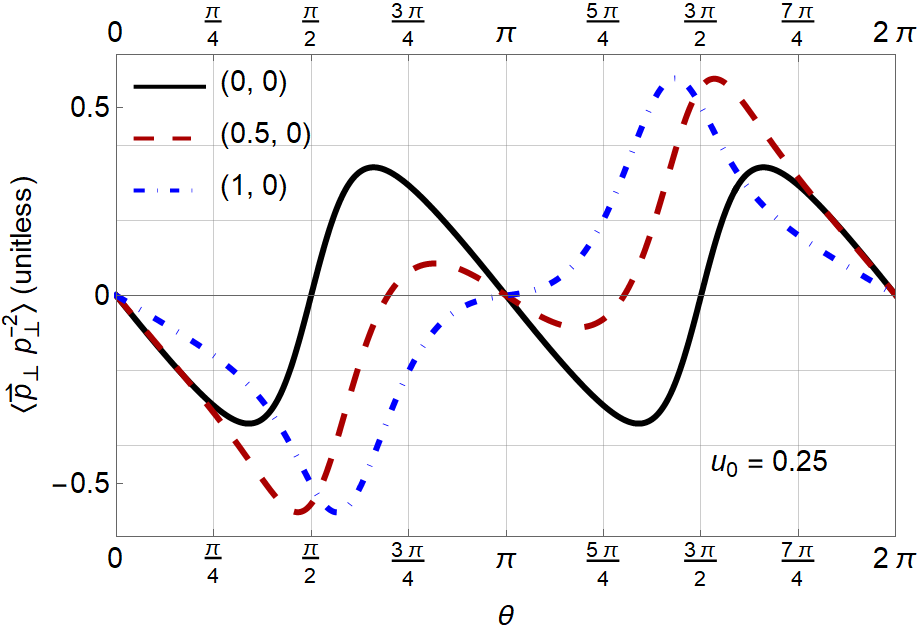}
\caption{Comparison of the angular profiles of the moments for jets produced at different points in the medium, using the same sample elliptical profile with $W=0.5$ and $H=1$.}
\label{f:Ellipse_Profiles}
\end{centering}
\end{figure}
%
%_____________________________________________________________

The simple picture of jet drift arising from jets produced at the exact center of the fireball is significantly modified as a function of the jet production point $(x_0 , y_0)$, as shown in Fig.~\ref{f:Ellipse_Profiles} due to a combination of path-length modulation and radial flow.  For production away from the center of the ellipse, the strength of the near-side attractor is only weakly attenuated as the production point moves out from $(0 , 0)$ toward $(W , 0)$.  The primary effect of moving the production point $(x_0 , 0)$ out toward the edge is the shrinking and disappearance of the away-side attractor at $\theta = \pi$.  As $x_0 \rightarrow W$, the repulsor directions initially at $\theta = \pi/2 , 3\pi/2$ move closer together, converging toward $\theta = \pi$.  This can be understood as a shrinking of the basin of attraction toward the away side $\theta = \pi$ and a growth of the basin of attraction toward the near side $\theta = 0$.

The angular dependence for general production points $(x_0 , y_0)$ which are off the major/minor axes is even more complicated, smearing the features of the angular profile and making interpretation difficult.  Given that the jet production point will vary event by event within a dataset, this raises the question of whether any of the simple geometric features seen here will survive statistical averaging over a fluctuating dataset.  To study this question in detail, we systematically introduce various sources of event-by-event fluctuations in Sec.~\ref{sec:fluct} below.

%||||||||||||||||||||||||||||||||||||||||||||||||||||||||||||||||||||||||||
%
\subsection{Event-By-Event Correlations to Elliptic Flow}
\label{sec:fluct}
%
%||||||||||||||||||||||||||||||||||||||||||||||||||||||||||||||||||||||||||

Based on the initial analysis of jet drift as a function of production point $(x_0 , y_0)$ for the fixed elliptical geometry in Sec.~\ref{sec:staticellipse}, we now want to pursue a statistical analysis of the jet drift moments over a dataset which incorporates various sources of event-by-event fluctuations.  The first we will consider is fluctuations in the jet production point $(x_0 , y_0)$, which we will implement using a phenomenologically-motivated proxy for binary collision weighting.

Unlike the soft particles making up the plasma, jets are initially produced in hard-scattering processes between partons from the two colliding nuclei.  The probability $P_\mathrm{hard}$ for such an initial hard scattering to occur is therefore proportional to the (two-dimensional) density of the first nucleus, denoted $T_A$, and to the density of the second nucleus, denoted $T_B$.  This product is referred to as the binary collision density $n_{BC}$, which should determine the probability distribution for where jets are produced in the resulting fireball:
\begin{align}   \label{e:BC1}
    P_\mathrm{hard} (x,y) \propto n_{BC} (x,y) = T_A (x,y) \: T_B (x,y) \: .
\end{align}
In our case, since the elliptical Gaussian toy model \eqref{e:ellipseT} is not built explicitly in terms of underlying nuclear densities $T_A , T_B$, we cannot employ the binary collision weighting \eqref{e:BC1} directly.  What we can do, however, is connect the two quantities indirectly through the entropy density $s(x,y)$. 

A flexible parameterization for the initial entropy density $s(x,y)$ in terms of the nuclear densities $T_A , T_B$ was implemented within the phenomenological Trento model \cite{Moreland:2014oya, Moreland:2018gsh} and later subjected to full-scale Bayesian analysis \cite{Bernhard:2016tnd} to determine what ``averaging procedure'' of the two nuclear profiles best predicted the initial state of a heavy-ion collision.  The Bayesian analysis found a significant preference for an initial entropy density proportional to the geometric mean,
\begin{align}   \label{e:BC2}
    s(x,y) \propto \sqrt{ T_A (x,y) \, T_B (x,y) }  \: .
\end{align}
From dimensional analysis, we know that the entropy density must be proportional to the cube of the temperature profile, $s \propto T^3$.  Using this, we can estimate the binary collision density $n_{BC}$ directly from our temperature profile \eqref{e:ellipseT} by combining Eqs.~\eqref{e:BC1} and \eqref{e:BC2}, obtaining
\begin{align}   \label{e:BC3}
    P_\mathrm{hard} (x,y) &\propto s^2 (x,y) \propto T^6 (x,y)
    \notag \\
    &\propto
    \exp\left[-\frac{x^2}{2 (\sigma_x/\sqrt{6})^2}\right] \, \exp\left[-\frac{y^2}{2 (\sigma_y/\sqrt{6})^2}\right]
\end{align}
in terms of the temperature.  Since the profile \eqref{e:ellipseT} is assumed to be Gaussian, raising it to the $6^\mathrm{th}$ power simply yields a binary collision density which is also Gaussian, with $\sigma_x$ and $\sigma_y$ scaled down by a factor of $\sqrt{6}$.

%_____________________________________________________________
\begin{figure}[!ht]
\begin{centering}
\includegraphics[width=0.48\textwidth]{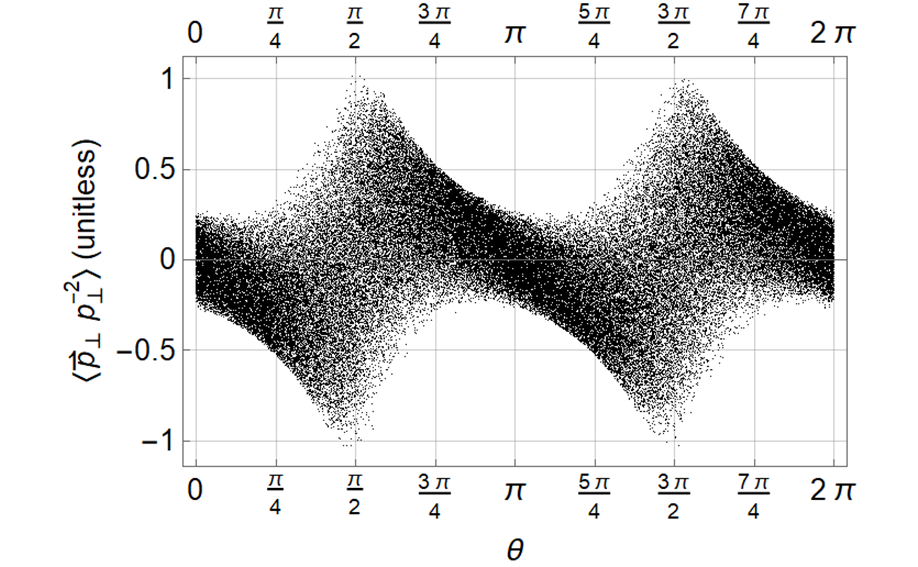}
\includegraphics[width=0.48\textwidth]{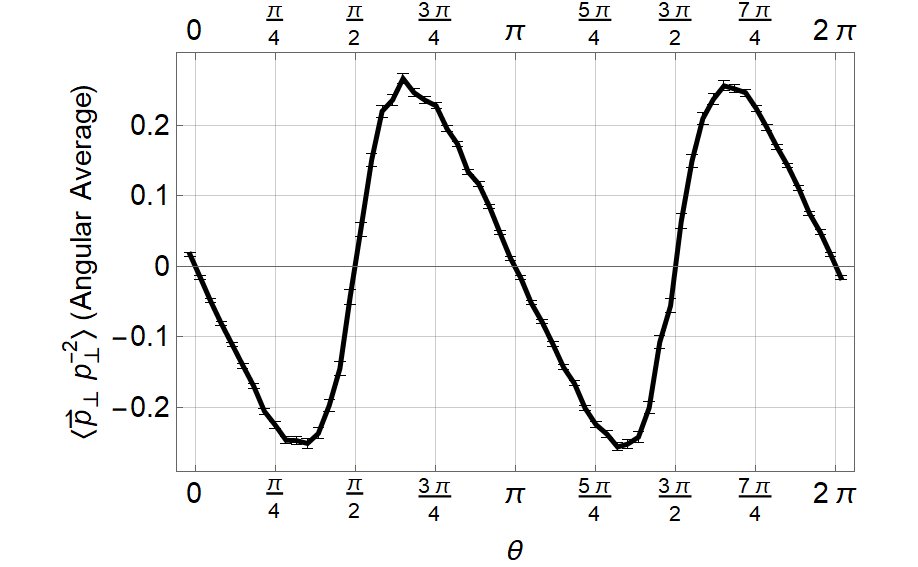}
\caption{Left: Scatter plot of jet drift moments for a dataset with fluctuating jet production points on a fixed geometry. Right: Jet drift moments binned by angle $\theta$ and averaged over jet production points with jackknife error estimation.
\label{f:FixGeom}
}
\end{centering}
\end{figure}
%_____________________________________________________________

In this way, we can construct a representative dataset with binary-collision-weighted jet production by sampling $(x_0, y_0)$ from a Gaussian of height and width $H / \sqrt{6}$ and $W / \sqrt{6}$, respectively, and sampling the initial jet direction $\theta$ from a uniform distribution $\theta \in [0,2\pi)$.  The results are shown Fig.~\ref{f:FixGeom}.  The scatter plot reflecting a sample of $10^5$ production points over a single fixed geometry exhibits a significant spread in the acoplanarity moments, but a remaining angular modulation is clearly still visible by eye.  We isolate that angular modulation by computing the mean value of the acoplanarity moment in each angular bin, with errors quantified using jackknife resampling.  The resulting plot of the mean acoplanarity moment $\langle \: \langle \tvec{p} p_\bot^{-2} \rangle \: \rangle_{(x_0, y_0)}$ as a function of angle bears a striking resemblance to the case of production from the center of the fireball seen in Fig.~\ref{f:EllipseSlice1}.  The quadropolar pattern of signs in each quadrant is preserved, and the locations of the zero crossings occur precisely at $\theta = 0, \pi/2, \pi, 3\pi/2$, reflecting the stability of the geometry coupling despite event-by-event fluctuations in the production point.  This suggests that the attractor/repulsor physics and its correlation to the elliptical geometry is quite robust against fluctuations.

%_____________________________________________________________
\begin{figure}[!ht]
\begin{centering}
\includegraphics[width=0.48\textwidth]{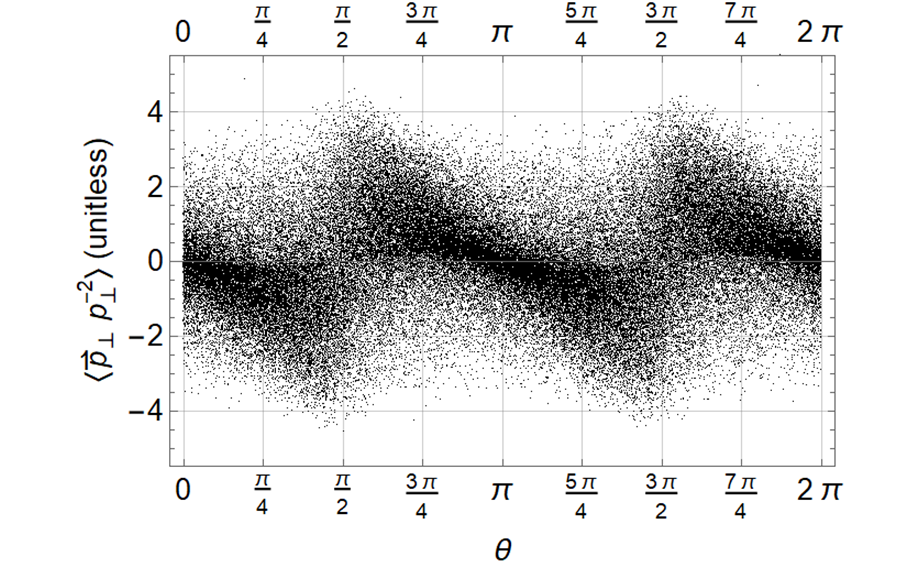}
\includegraphics[width=0.48\textwidth]{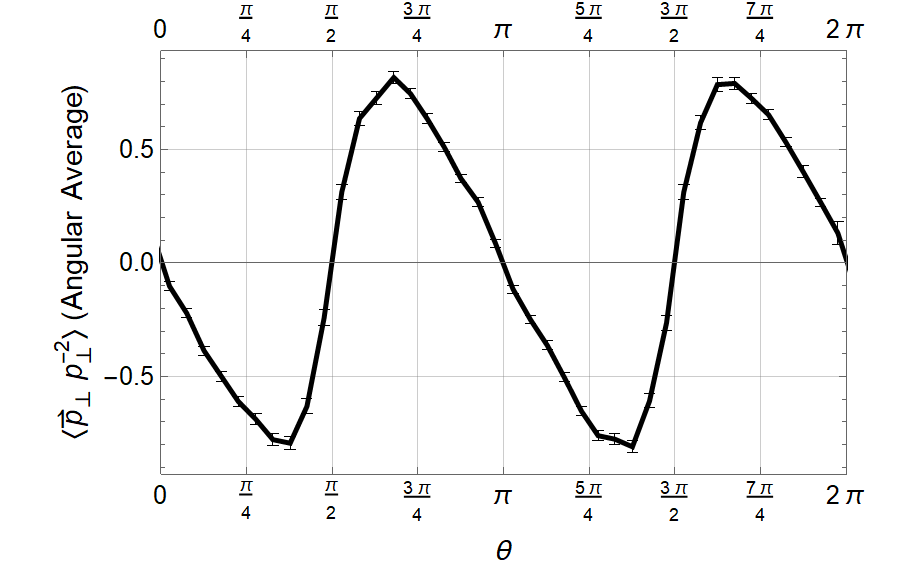}
\caption{Left: Scatter plot of jet drift moments for a dataset with fluctuating jet production points on fluctuating impact parameter. Right: Jet drift moments binned by angle $\theta$ and averaged over jet production points and impact parameters $b$ with jackknife error estimation.
\label{f:FluctGeom}
}
\end{centering}
\end{figure}
%_____________________________________________________________

That continues to be the case when we introduce fluctuations in the underlying geometry itself.  By sampling the impact parameter $b \in [0, 2 R)$ from a uniform distribution, it is straightforward to compute an event-by-event temperature profile using \eq{e:HW1} which reflects the overlap probability of two colliding beams of hard-sphere nuclei.  Constructing a dataset with both fluctuating geometry and production point along the same lines leads to the plots shown in Fig.~\ref{f:FluctGeom}.  The addition of geometry fluctuations leads to a greater spread of points in the scatter plot, but again the geometry-driven $\sim \cos2\theta$ modulation persists.

%_____________________________________________________________
\begin{figure}[!ht]
\begin{centering}
\includegraphics[width=0.32\textwidth]{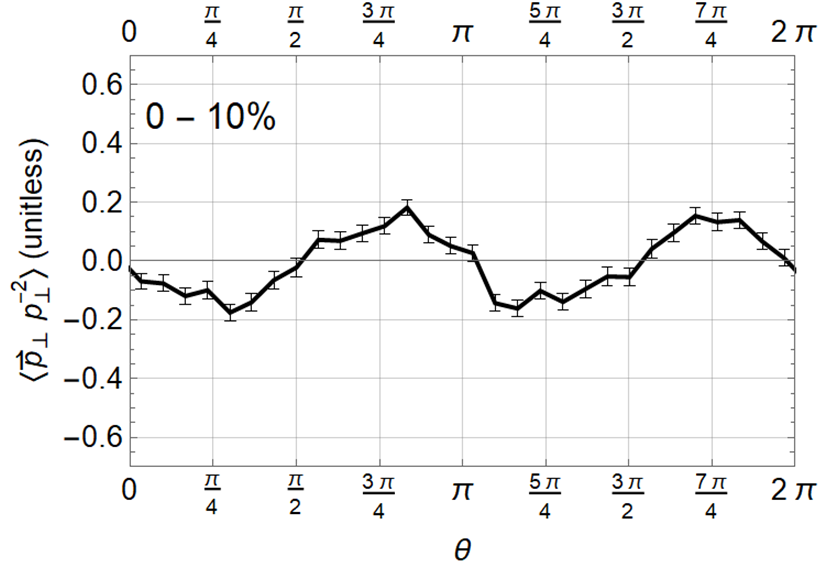}
\includegraphics[width=0.32\textwidth]{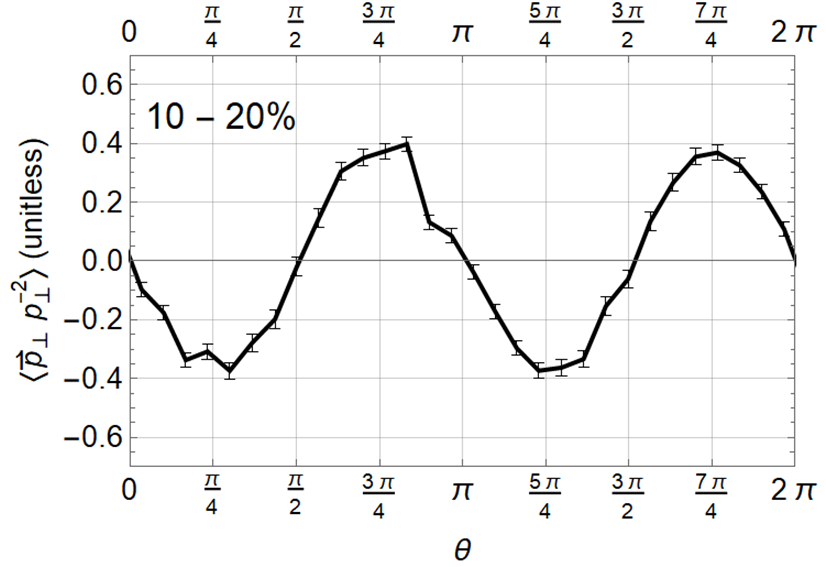}
\includegraphics[width=0.32\textwidth]{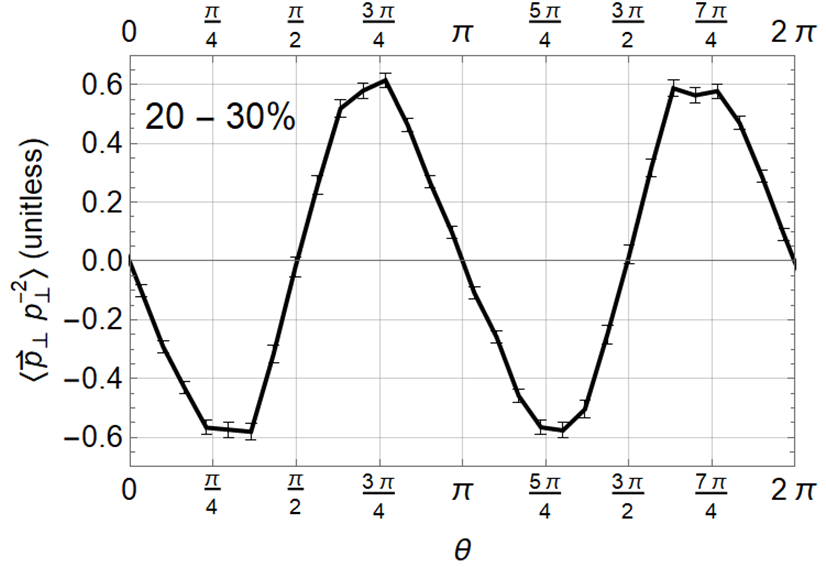}
\caption{Moments averaged over production point and geometry within corresponding centrality bins. Note that the magnitude of our effect increases with centrality.
\label{f:CentBins}
}
\end{centering}
\end{figure}
%_____________________________________________________________

The inclusion of event-by-event geometry fluctuations allows us to go even further by performing centrality binning.  Given an initial entropy density profile $s \propto T^3$, we can integrate to obtain the total entropy deposited in a given event, which is in turn roughly proportional to the multiplicity $N_\mathrm{soft}$ of soft particles detected in the final state:
\begin{align}
    N_\mathrm{soft} &\propto \int dx dy \, T^3 (x,y) \propto H W    \: ,
\end{align}
which is just the area of the initial state.  Thus we can partition the dataset obtained by uniform sampling of the impact parameter $b$ into ranked quantiles based on the multiplicity $N_\mathrm{soft} \propto H W$ to form centrality classes, just like in the experimental analysis.
Breaking the dataset into centrality classes and performing the same analysis in each bin, we obtain the plots shown in Fig.~\ref{f:CentBins}.  The same $\cos2\theta$ modulation again persists in each centrality class, with a magnitude that grows with increasing centrality.\footnote{While the absolute scale of the jet drift moments shown here is arbitrary, the relative change as a function of centrality is meaningful.}  

%_____________________________________________________________
\begin{figure}[ht]
\begin{centering}
\includegraphics[width=0.48\textwidth]{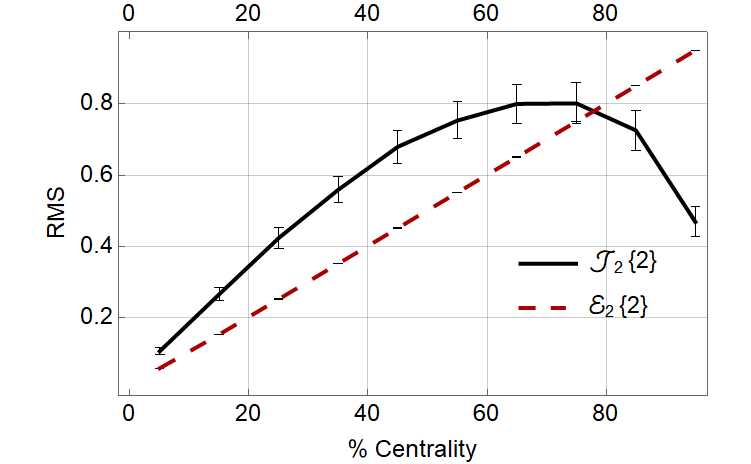}
\caption{Centrality dependence of the RMS jet drift $\mathcal{J}_2 \{2\}$ from \eq{e:jetcumulant} and the RMS ellipticity $\varepsilon_2 \{2\}$.  As seen, the two quantities are highly correlated.
\label{f:RMScorrels}
}
\end{centering}
\end{figure}
%_____________________________________________________________

We can single out the strength of this modulation as a function of centrality by computing the root-mean-square value of the jet moment as a function of theta, which we denote as 
\begin{align}   \label{e:jetcumulant}
    \mathcal{J}_2 \{2\} \equiv
    \left\langle\:
    \left\langle
        \tvec{p} p_\PrpXY^k
    \right\rangle_{x_0,  y_0, b}^2
    \: \right\rangle_\theta^{1/2} \: ,
\end{align}
in analogy with the soft sector cumulants like $v_2 \{2\}$.  The reason for the growth of the jet drift effect with increasing centrality is presumably due to the increase in the ellipticity of the event, which can be quantified through the complex eccentricity vector $\bm{\mathcal{E}}_2$.  The eccentricity, with magnitude $\varepsilon_2$ and event-plane angle $\psi_2$ is defined by
\begin{align}   \label{e:ecc2_1}
    \bm{\mathcal{E}}_2 \equiv \varepsilon_2 \, e^{2 i \psi_2} &\equiv
    -
    \frac{
        \int r dr d\theta \, r^2 \, e^{2 i \theta} \, s(r, \theta)
    }{
        \int r dr d\theta \, r^2 \, s(r, \theta)
    }
    =
    -
    \frac{
        \int dx dy \, (x + i y)^2 \, s(x,y)
    }{
        \int dx dy \, (x^2 + y^2) \, s(x,y)
    }
    \notag \\ &=
    \frac{H^2 - W^2}{H^2 + W^2} \: ,
\end{align}
where in the last step we have evaluated the integrals for the Gaussian profile \eqref{e:ellipseT}.  We plot in Fig.~\ref{f:RMScorrels} both the RMS acoplanarity moment $\mathcal{J}_2 \{2\}$ in a given centrality bin and the RMS eccentricity, known as the second cumulant $\varepsilon_2 \{2\} \equiv \langle \varepsilon_2^2 \rangle^{1/2}$.  In this simple model the eccentricity $\varepsilon_2 \{2\}$ grows linearly with centrality, and across most of the centrality region so does the magnitude of the RMS asymmetric acoplanarity moment $\mathcal{J}_2 \{2\}$.  These two quantities are highly correlated with each other, reflecting the strength of the elliptical geometry on the direction of the jet drift.  Moreover, because of the extremely low viscosity of the quark-gluon plasma, the eccentricity $\bm{\mathcal{E}}_2$ of the initial state is a strong estimator of the final-state elliptic flow $\bm{\mathcal{V}}_2$ of the measured final-state.  Thus a correlation between the acoplanarity moments $\langle \tvec{p} p_\PrpXY^k \rangle$ and $\bm{\mathcal{E}}_2$ may be directly translated into a correlation between jet drift and the elliptic flow $\bm{\mathcal{V}}_2$ which could be measured experimentally.

The last layer of event-by-event fluctuations we can include are associated with event-by-event fluctuations in the direction of the event plane $\psi_2$ and the associated direction of the acoplanarity asymmetry.  Thus far we have allowed the impact parameter $b$ to fluctuate from event to event, but we have kept the orientation of the ellipse fixed such that the impact vector $\vec{b}$ always pointed along the $x$-axis, giving $\psi_2 = 0$.  Typically in experimental datasets, however, it is not possible to bin the data in such a way that the event plane angle is fixed -- with a notable exception being the capabilities enabled by the STAR Event Plane Detector \cite{Adams:2019fpo}.  Using the ability to bin heavy-ion events by centrality and reaction plane angle, it may be possible to compare the correlation between the jet drift moments and geometry at the level of Figs.~\ref{f:FixGeom} and \ref{f:RMScorrels}.  But in most cases, experimental datasets can be binned by centrality (correlated with the magnitude of $\vec{b}$), but the direction of the reaction plane $\psi_2$ will fluctuate uncontrollably from event to event.  

To capture the directional correlation of the weighted jet acoplanarity $(\tvec{p} p_\PrpXY^k)$ with the fluctuating geometry, we must sample \textit{both} the reaction plane angle $\psi_2$ and the direction of $\tvec{p}$ on an event-by-event basis.  While it is trivial to randomly sample $\psi_2 \in [0,2\pi)$ from a uniform distribution, this also means that we cannot directly evaluate the acoplanarity moments using \eq{e:elpmom1}; we must instead perform Monte Carlo sampling of the underlying distribution \eqref{e:elpdistr1}.  This introduces two additional subtleties which we must address now.  First, even if the photon (and therefore the initial jet) lies in the $xy$-plane at mid rapidity, the transverse momentum $\vec{p}_T$ acquired from scattering in the medium may scatter the jet out of that plane if $p_z \neq 0$.  This broadening of the rapidity distribution contains important physical information, but it is beyond the scope of the $2+1$D description of the initial state we are considering here.  The boost-invariant approximation we use here can be thought of as assuming that the same $\gamma + \mathrm{jet}$ production process occurs identically at each rapidity slice, so that for every jet scattered out of the $xy$-plane with $p_z \neq 0$, another jet is scattered \textit{into} the $xy$-plane from a different initial rapidity.  Moreover, for a flow field $\vec{u}$ which is boost invariant and limited to the $xy$-plane, the preferred direction of $\tvec{p}$ will always lie in the $xy$-plane; thus this out-of-plane broadening serves only to dilute the statistical significance of the directional correlations, not introduce correlations of its own.  For these reasons, we will perform only a sampling of the component $(p_\bot \hat{e}_\PrpXY)$ lying in the $xy$-plane, while setting $p_z = 0$ by hand.  

The other complication which arises from needing to consider an event-by-event sampling of the distribution \eqref{e:elpdistr1} is the pathological behavior of the \textit{symmetric part} of the distribution when $p_\PrpXY \rightarrow 0$.  Unlike the antisymmetric term responsible for the acoplanarity moments \eqref{e:elpmom1} which explicitly vanishes if $p_\PrpXY \rightarrow 0$, the symmetric part of the distribution is peaked at $p_\PrpXY = 0$ with a height controlled by the Debye mass $\mu$.  This reliance on the Debye mass for a regulator at $p_\PrpXY = 0$ leads to the pathological behavior if we allow $\mu = g T$ to decrease to zero in the periphery of the fireball.  We can clearly see the origin of the pathological behavior if we consider a jet traveling along the $x$-axis with $p_\PrpXY = 0$, for which the line integral
\begin{align}   \label{e:mu0problem}
    \frac{dN^{(1)}}{d^2 p_\PrpXY} \overset{p_\PrpXY = 0}{\sim }
    \int \frac{dt}{\lambda(t)} \frac{g^2 T^2 (t)}{(0 + g^2 T^2 (t))^2}
    \sim \int \frac{dt}{T(t)} 
    \sim \int dt \, e^{+ t^2 / 2 \sigma_x^2}
\end{align}
blows up exponentially at large distances, leading to large / divergent contributions to the height of the peak at $p_\PrpXY = 0$ coming from far outside the fireball itself.  Mathematically, this unphysical behavior occurs because even though the density of the medium is falling rapidly to zero, the regulator $\mu = g T$ which limits the range of gluon exchange with the medium to distances $d \lesssim 1/\mu$ is also going to zero.  This allows infinitely soft, infinitely long-ranged gluon exchange between the jet and the medium, generating uncontrolled contributions to the peak at $p_\PrpXY = 0$.  While one could attempt to regulate this unphysical behavior in a variety of ways (such as a distance or temperature cutoff), most choices will lead to an exponentially-sensitive dependence on the chosen cutoff because of the scaling illustrated in \eq{e:mu0problem}.  Instead, we will regulate this divergence by introducing a temperature-independent contribution $\mu_0$ to the Debye mass via
\begin{align}
    \mu^2 (t) \rightarrow g^2 T^2 (t) + \mu_0^2 \: ,
\end{align}
with $\mu_0$ some nonperturbative screening scale.  Not only is the height of the $p_\PrpXY = 0$ peak much more weakly-dependent on the choice of $\mu_0$ in this regularization scheme; the assignment of a nonperturbatively-generated scale $\mu_0$ also mirrors the actual onset of confinement physics at temperature scales so low that the perturbative relation $\mu = g T$ should break down.  In this sense the cutoff $\mu_0$ we assign here can be seen to play a role analogous to that of the nonperturbative magnetic mass $\mu_M$ as written in the generalized hard thermal loop cross section given in \eq{e:HTLplus}.

%--------------------------------------------------------------------------
\begin{figure}
\begin{centering}
\includegraphics[width=0.48\textwidth]{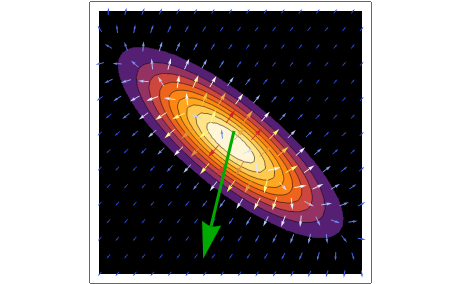}
\includegraphics[width=0.48\textwidth]{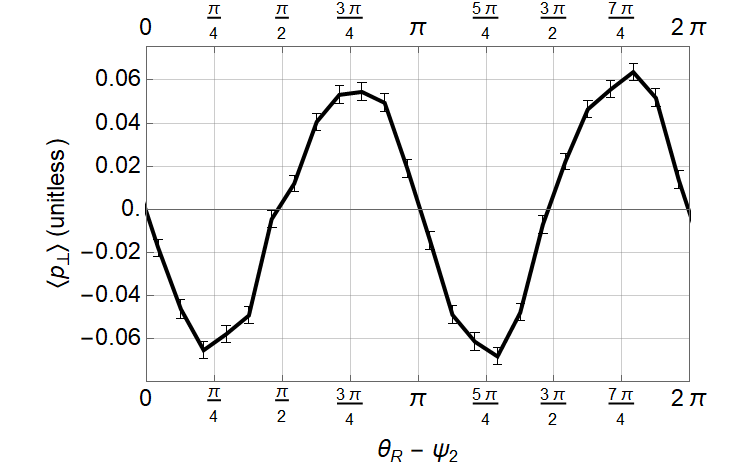}
\caption{Event-by-event sampling of the reaction plane $\psi_2$ and acoplanarity $p_\PrpXY$. Left panel: the plasma profile \eqref{e:elpdistr2} for a particular event, together with jet production point and angle depicted (green arrow).  Right panel: the resulting angular correlation with the reaction plane after sampling individual events as in the left panel many times.} 
\label{f:deflsample}
\end{centering}
\end{figure}
%--------------------------------------------------------------------------

With both of these modifications to the distribution \eqref{e:elpdistr1}, we perform a Monte Carlo rejection sample of the distribution
\begin{align}   \label{e:elpdistr2}
&\frac{dN^{(1)}}{d^2 p_T} 
    = 
    \frac{dN^{(1)}}{d p_\bot \, dp_z} 
    \notag \\ &=
    \delta (p_z) \, \frac{g^2}{\pi \kappa} \int dt \, \frac{T^3 (t)}{((\tvec{p} \cdot \hat{e}_\PrpXY)^2 + g^2 T^2(t) + \mu_0^2)^2} \:\notag
    \\
    & \hspace{1in} \times \left[ 1 + 
    \frac{\vec{p} \cdot \hat{e}_\PrpXY}{E} \:
    \frac{\vec{u} (t) \cdot \hat{e}_\PrpXY}{1 - \vec{u} (t) \cdot \hat{e}_\parallel} \:
    \frac{6 (\tvec{p} \cdot \hat{e}_\PrpXY)^2 + 4 g^2 T^2 (t) }{(\tvec{p} \cdot \hat{e}_\PrpXY)^2 + g^2 T^2 (t)} \right] \: ,
\end{align}
resulting in the angular distribution shown in Fig.~\ref{f:deflsample}.  As illustrated, the elliptical Gaussian profile \eqref{e:ellipseT} is now rotated by a random angle $\psi_2$ with respect to the coordinate axes.  As a result, the mean acoplanarity $\langle \tvec{p} \rangle$ is generally uncorrelated with the coordinate axes, but one recovers the same quadrupolar modulation seen previously when expressed relative to the reaction plane as a function of $\theta - \psi_2$.

The persistence of the jet drift / geometry coupling despite the addition of all of the event-by-event fluctuations possible within the simple toy model of Sec.~\ref{sec:model} speaks to the robustness of the effect.  While there are many details of this simple model which are expected to be modified in realistic hydrodynamic simulations, the preferred direction of the elliptic flow -- and its inevitable consequences on the pattern of jet drift -- is not one of them.  With fully fluctuating event-by-event variables, we can now formulate an experimental observable which expresses this fundamental jet drift / geometry correlation.  We propose that the physics of jet drift / geometry coupling can be measured by studying the correlation
\begin{align*}
    \left\langle \tvec{p} \cdot \bm{\mathcal{V}}_2 \right\rangle =
    \left\langle (\tvec{p} \cdot \hat{e}_\PrpXY) \, v_2 \, e^{2 i (\theta - \psi_2)} \right\rangle
\end{align*}
between the jet acoplanarity and the elliptic flow of soft particles.  More generally, one can decompose the measured angular distribution of jet acoplanarity as in Fig.~\ref{f:deflsample} into various Fourier harmonics, exactly as in the anisotropic flow $\bm{\mathcal{V}}_n$ in the soft sector \cite{Luzum:2013yya}.  Then the relevant correlation can be expressed as an appropriately normalized Pearson correlation coefficient
\begin{align}   \label{e:obs1}
    \mathcal{C}[\tvec{p} , \bm{\mathcal{V}}_2] &\equiv
    \frac{
        \left\langle (\tvec{p} \cdot \hat{e}_\PrpXY) \, v_2 \, e^{2 i (\theta - \psi_2)} \right\rangle
    }{
        \mathcal{J}_2 \{2\} \:
        v_2 \{2\}
    }   
    \approx
    \frac{
        \left\langle (\tvec{p} \cdot \hat{e}_\PrpXY) \, \varepsilon_2 \, e^{2 i (\theta - \psi_2)} \right\rangle
    }{
        \mathcal{J}_2 \{2\} \:
        \varepsilon_2 \{2\}
    }
    = \mathcal{C}[\tvec{p} , \bm{\mathcal{E}}_2] \: .
\end{align}
In the last step, we have related the elliptic flow $\bm{\mathcal{V}}_2$ of the final state to the ellipticity $\bm{\mathcal{E}}_2$ of the initial state through approximate linear response $\bm{\mathcal{V}}_2 \propto \bm{\mathcal{E}}_2$.  Such linear response is known to be a good approximation for a variety of collision systems, especially in more central collisions, with higher-order nonlinear response playing a smaller role \cite{Rao:2019vgy, Noronha-Hostler:2015dbi}.

%--------------------------------------------------------------------------
\begin{figure}
\begin{centering}
\includegraphics[width=0.49\textwidth]{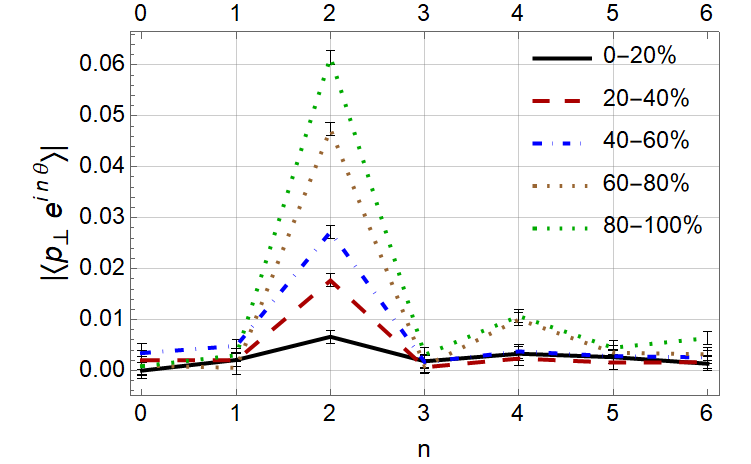}
\includegraphics[width=0.49\textwidth]{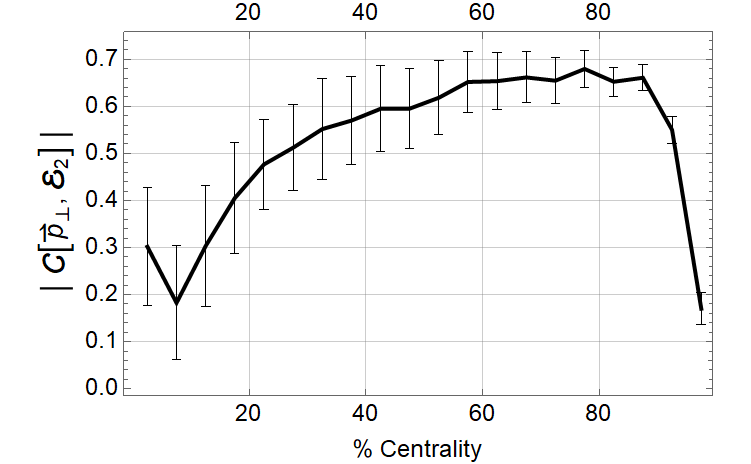}
\caption{Left: Harmonics of $|\langle  p_\perp \rangle|$ as a function of centrality class,  }
\label{f:newobs}
\end{centering}
\end{figure}
%--------------------------------------------------------------------------

The Fourier harmonics obtained by decomposing the angular distribution shown in Fig.~\ref{f:deflsample} and the properly-normalized Pearson coefficient \eqref{e:obs1} are shown in Fig.~\ref{f:newobs}.  The Fourier harmonics shown in the left panel of Fig.~\ref{f:deflsample} confirm the features of the angular distribution of Fig.~\ref{f:deflsample} seen previously, notably the very strong $\cos 2(\theta - \psi_2)$ modulation which dominates all other harmonics.  The odd harmonics $n = 1, 3, 5$ are consistent with zero as expected due to the exact mirror symmetry of the optical Glauber model employed here; while these harmonics need not be exactly zero in realistic events, they are expected to be much smaller than the strong $n=2$ signal which is present already at the mean-field level.  The strength of the $n=2$ elliptic modulation of the jet drift grows with increasing centrality, as seen previously in Fig.~\ref{f:RMScorrels}.  Interestingly, there is also a nontrivial growth of the higher-order even harmonics $n = 4, 6$ with centrality as well which may be responsible for producing the different slopes at the attractor/repulsor zero crossings as seen by eye in Fig.~\ref{f:EllipseSlice1} and Eq.~\eqref{e:slopes}.  The properly-normalized Pearson coefficient \eqref{e:obs1} shown in the right panel of Fig.~\ref{f:newobs} is sizeable and nonzero, growing as a function of centrality due to the increasing ellipticity of the fireball.  This growth continues out to very peripheral events, where a mean-field description at the level of the simple model \eqref{e:ellipseT} should break down.  While this correlation may break down at very peripheral collisions, the sizeable correlation and growth with centrality out to around $\sim 60\%$ centrality are expected to persist in more realistic simulations.

%%%%%%%%%%%%%%%%%%%%%%%%%%%%%%%%%%%%%%%%%%%%%%%%%%%%%%%%%%%%%%%%%%%%%%%%%%%
%
\section{Conclusions}
\label{sec:concl}
%
%%%%%%%%%%%%%%%%%%%%%%%%%%%%%%%%%%%%%%%%%%%%%%%%%%%%%%%%%%%%%%%%%%%%%%%%%%%

In this paper, we have performed a systematic analysis of the physics of jet drift due to collective flow in heavy-ion collisions.  By starting in Sec.~\ref{sec:slab} with the simplest possible toy model, a constant slab of flowing plasma, and gradually introducing in Sec.~\ref{sec:geometry} the effects of finite geometries, elliptical deformation, and event-by-event fluctuations, we have transparently pinpointed the underlying physics of jet drift and its correlation to the bulk geometry of heavy-ion collisions.  We find quite generally that the jet is dragged in the direction of the fluid flow, which serves as an attractor of the long-time evolution in an infinite medium.  In finite elliptical media, this drift effect leads to the preferential deflection of jets in the direction of the flow, which is in turn strongly correlated with the ellipticity of the initial geometry.  We have proposed in \eq{e:obs1} a new observable -- the correlation coefficient of the $\gamma + \: \mathrm{jet}$ acoplanarity with the elliptic flow of soft particles -- which can be studied experimentally to quantify this correlation.  From elementary line integrals over fixed backgrounds to fully-fluctuating event-by-event simulations within a simplified optical Glauber model, we find that this correlation is quite robust at all levels.  By studying these key results within two extremal models (Gyulassy-Wang versus Hard Thermal Loop), we have further established in Appendix~\ref{sec:HTL} that these general conclusions are largely insensitive to the underlying microscopic model.  Thus we conclude that jet drift as a physical phenomenon and its natural coupling to the geometry of a heavy-ion collision are robust and may be observable in real experiments through measurements of the correlation \eqref{e:obs1}.

While the series of simplified models we have considered here are far from realistic descriptions of heavy-ion collisions, they do capture the essential features of elliptic geometry and flow, centrality dependence, and event-by-event fluctuations.  The purpose of studying these toy models was to identify the most important features of jets' response to collective flow in heavy-ion collisions and design observables which are sensitive to these features.  The limitation of these simple models is that it is difficult to estimate the typical size of the angular deflection and whether it will be measurable given the limitations of finite statistics, acceptance, and background subtraction.  Back-of-the-envelope parametric estimates based on the constant slab shown in Fig.~\ref{f:DefVE} suggest that the typical angular deflection for jets with energies in the tens of GeV could be on the order of a few degrees.  If these estimates continue to hold under more realistic simulations, then jet drift and its correlation to elliptic flow may well be measurable either at sPHENIX or at the lower end of jet acceptance at the LHC.  While one could continue the program begun here of gradually incorporating the features of realistic heavy-ion collisions one step at a time, at this point there is no obstacle to immediately studying jet drift in full-fledged $2+1$D viscous relativistic hydrodynamics.  This extension is our top priority for follow-up work, in order to make realistic predictions of the absolute size of the jet deflection to determine whether it will be observable in experiment.  Since it takes time to accumulate large radial flow in heavy-ion collisions, it may be important to incorporate a realistic treatment of jets interacting with the hadron gas phase \cite{Dorau:2019ozd} at the latest times in the QGP evolution.

The fundamental physics of jet drift we study here, along with other asymmetric measures of jet-medium coupling, can also be extended to a number of other physical mechanisms and processes.  Photon-jet events as described here are a convenient theoretical testbed since the photon provides a clean probe of the initial jet direction, but they are relatively rare and thus statistics-limited.  Generalizing this approach to the phenomenology dijets or dihadrons, including the incorporation of appropriate jet or fragmentation functions will be important to maximize the ability to statistically distinguish the asymmetric drift effect from the symmetric background.  Another key phenomenological question will be the impact of medium flow on the pattern of soft gluon radiation which is predominantly responsible for jet energy loss.  While a more complex phenomenon, the asymmetry of soft gluon radiation was also a major result derived in Ref.~\cite{Sadofyev:2021ohn} with many similarities to the jet drift effect we have studied here.  Tellingly, the asymmetric part of the medium-induced radiation spectrum scales in a familiar way with the medium flow:
\begin{align}   \label{e:rad1}
    E_\gamma E \omega \: \frac{dN^{(1)}}{d^3 p_\gamma \, d^3 p \, d^3 k} \sim
    \int\frac{dt}{\lambda} \frac{\tvec{u} (t) \cdot \tvec{k}}{1 - u_\parallel (t)}
    \sim
    \tvec{k} \cdot \left\langle \frac{\tvec{p}}{p_\PrpXY^2} \right\rangle \: ,
\end{align}
where the radiated soft gluon carries energy $\omega$ and transverse momentum $\tvec{k}$.  The line integral appearing in \eq{e:rad1} which controls the asymmetry of soft gluon radiation is exactly of the form \eqref{e:vecmoment2} describing the jet drift moments; in particular, it is exactly the moment $\langle \tvec{p} / p_\PrpXY^2 \rangle$ for $k = -2$.  Thus we anticipate that many of the general features we observe here for jet drift, including its significant correlation to the elliptic flow, will also hold true when applied to the asymmetric pattern of soft gluon radiation emitted by the jet.

Finally, we note that in this paper we have emphasized the coupling of jet drift to the pattern of radial flow which develops at mid-rapidity in heavy-ion collisions.  This is only one possible place to look for signatures of jets being dragged in the direction of the flowing plasma.  The other natural consideration is the role of the rotational motion of the QGP induced by the deposition of tremendous amounts of orbital angular momentum into the plasma by the colliding nuclei.  Signatures of the ``most vortical fluid in nature'' have already been seen in the conversion of some part of this orbital angular momentum to the spin of produced hyperons \cite{STAR:2017ckg}, and the coupling of jets to this tremendous rotational motion could also leave observable experimental signatures.  Given all of the open questions and extensions our work raises, it is clear that we have only begun to scratch the surface of what can be learned from jet drift physics in heavy-ion collisions.  The continued development of new observables which are insensitive to traditional backgrounds and may permit the extraction of novel information about the QGP gives renewed promise to the long-sought program of mature jet tomography.

%%%%%%%%%%%%%%%%%%%%%%%%%%%%%%%%%%%%%%%%%%%%%%%%%%%%%%%%%%%%%%%%%%%%%%%%%%%
%
\section*{Acknowledgments}
%
%%%%%%%%%%%%%%%%%%%%%%%%%%%%%%%%%%%%%%%%%%%%%%%%%%%%%%%%%%%%%%%%%%%%%%%%%%%

The authors would like to thank
Travis Dore,
Christine Nattrass and her research group at the University of Tennessee, Knoxville,
Jorge Noronha,
Jacquelyn Noronha-Hostler,
Andrey Sadofyev,
Ivan Vitev, and
Boram Yoon
for insightful discussions.  This work is supported by a startup grant from New Mexico State University and by the generous donation of computing resources from Jeremy Tanner and Kthoris Scientific Computing Services.

\appendix

%%%%%%%%%%%%%%%%%%%%%%%%%%%%%%%%%%%%%%%%%%%%%%%%%%%%%%%%%%%%%%%%%%%%%%%%%%%
%
\section{Real-Time Evolution in the Constant Slab}
\label{sec:trajectory}
%
%%%%%%%%%%%%%%%%%%%%%%%%%%%%%%%%%%%%%%%%%%%%%%%%%%%%%%%%%%%%%%%%%%%%%%%%%%%

%--------------------------------------------------------------------------
\begin{figure}
\begin{centering}
\includegraphics[width=0.49\textwidth]{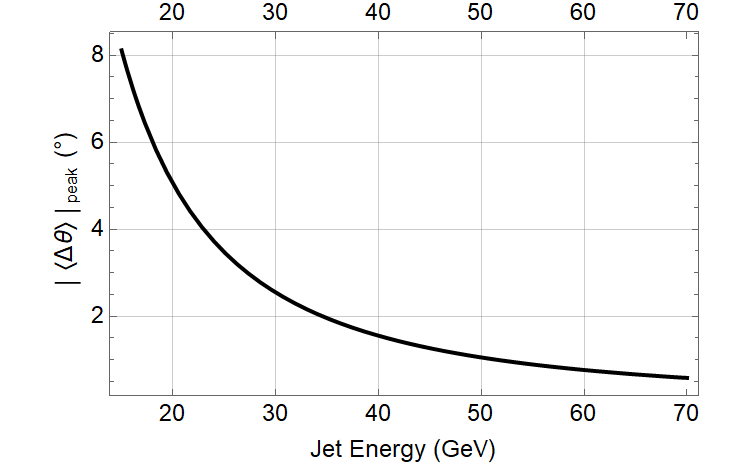}
\caption{Graph illustrating the effect of different jet energies on maximum absolute angular deflection.
\label{f:DefVE}
}
\end{centering}
\end{figure}
%--------------------------------------------------------------------------

Alternative to the moment analysis considered above, we can also extract significant information about the jet trajectories directly from the angular deflection corresponding to the $k=0$ moment in \eq{e:meantheta1}.  One immediate use is to estimate the typical size of the angular deflection for jets of a given energy; taking the geometry to be a constant slab and setting the fluid direction $\alpha = 0$ for simplicity, we have
\begin{align} \label{e:deflection}
    \langle \Delta \theta \rangle = -3 \frac{L}{\lambda} \frac{u\sin{\theta}}{(1-u\cos{\theta})} \frac{\mu^2}{E^2} \ln{\frac{E}{\mu}} \, .
\end{align}
As seen previously in Fig.~\ref{f:MA1}, the maximum deflection occurs at the position given by \eq{e:thetapeak}, giving
\begin{align} \label{e:deflection2}
    \left| \langle \Delta \theta \rangle \right|_\mathrm{peak}  = 3 \frac{L}{\lambda} \frac{u}{\sqrt{1 - u^2}} \frac{\mu^2}{E^2} \ln{\frac{E}{\mu}} \, .
\end{align}
Choosing ballpark values for the parameters $L / \lambda = 4$, $u = 0.7$, and $\mu = g T = 2 (500 \, \mathrm{MeV}) = 1 \, \mathrm{GeV}$ \cite{Vitev:2008vk}, we plot $\left| \langle \Delta \theta \rangle \right|_\mathrm{peak}$ as a function of jet energy in Fig. \ref{f:DefVE}.  As is clear from the graph, 40-60 GeV jets can possibly have deviations around one degree, with less energetic jets having even larger deflection.  Based on these preliminary estimates, it seems reasonable that the jet drift effect may be measurable in jets produced at sPHENIX or even at the lower end of acceptance at the LHC.

It is also instructive to transform the line integral from \eq{e:meantheta1} into a differential equation by moving the infinitesimal $dt$ to the left-hand side of the equation, writing
\begin{align}   \label{e: jetdiffeq0}
    \frac{d \langle \Delta \theta \rangle }{dt}  &=
    \frac{3}{\lambda(t)} \frac{u_\PrpXY (t)}{1 - u_\parallel (t)} \,
    \frac{\mu^2 (t)}{E^2} \, \ln\frac{E}{\mu(t)} \, .
\end{align}
This expression was originally derived with reference to the original direction $\hat{e}_\parallel$ of the jet as identified from the recoil photon and used to calculate the expected deflection of the jet distribution.  But written this way, we can interpret it more differentially: if we advance the position and direction of the jet according to \eq{e: jetdiffeq0} over one timestep $dt$, then recompute the deflection of the jet at the next timestep, we effectively construct the ``trajectory of a mean jet'' through the medium.  While this deterministic time evolution of the mean jet trajectory is simplistic and semiclassical, it is nonetheless highly insightful.  Taking the direction of the fluid flow to be $\alpha = 0$ for simplicity, we express $u_\PrpXY (t)$ and $u_\parallel (t)$ with respect to the average jet direction $\langle \theta \rangle$ to recast the line integral \eqref{e:meantheta1} as a differential equation for the mean trajectory $\langle \theta (t) \rangle$:
\begin{align} \label{e: jetdiffeq}
    \frac{d\langle\theta\rangle}{dt}=\frac{-3}{\lambda}\frac{u\sin{\langle\theta\rangle}}{(1-u\cos{\langle\theta\rangle})}\frac{\mu^2}{E^2}\ln{\frac{E}{\mu}} \, .
\end{align}
%

%--------------------------------------------------------------------------
\begin{figure}
\begin{centering}
\includegraphics[width=0.49\textwidth]{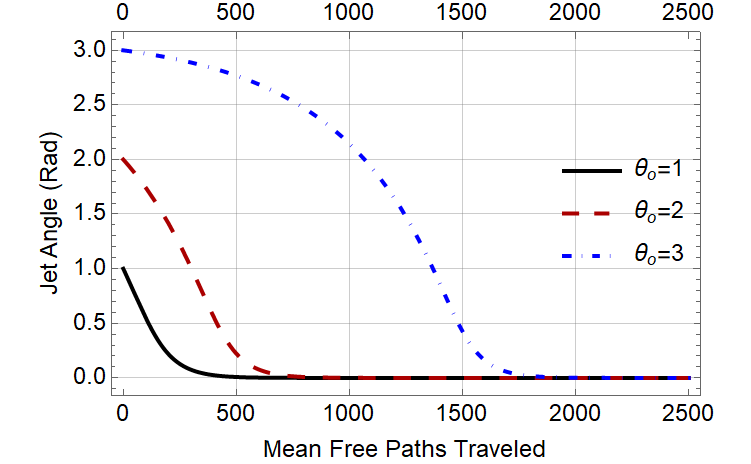}
\caption{Jet angle vs. path length for different $\theta_0$, with $E=50$ GeV, $\mu=1$ GeV, and fluid speed $u=0.7$.
\label{f:DEsol1}
}
\end{centering}
\end{figure}
%--------------------------------------------------------------------------

Eq.~\eqref{e: jetdiffeq} is solvable via separation of variables and yields the following relation when integrated from $\langle\theta\rangle=\theta_0\to\theta$, with $\theta_0$ being the start angle and $\theta$ being the average final angle, and from $t=0\to \Delta t$, the total path length:
\begin{align} \label{e: jetdiffeqsol}
    \ln{\left[\left(\frac{\sin{\theta_0}}{\sin{\theta}}\right)^{\frac{1}{u}+1}\left(\frac{1-\cos{\theta}}{1-\cos{\theta_0}}\right)^{\frac{1}{u}}\right]}=-\left(\frac{3}{\lambda}\frac{\mu^2}{E^2}\ln{\frac{E}{\mu}}\right)\Delta t
\end{align}
Fig. \ref{f:DEsol1} shows the solution $\langle \theta(t) \rangle$ for different initial conditions $\theta_0$. Since the fluid velocity direction has been set to $\alpha=0$, these curves show that, given enough path length, the jets will asymptotically align themselves with the direction of the fluid flow.  In this sense, the direction $\vec{u}$ of the fluid flow is a true attractor of the jet trajectory as a stable equilibrium solution of the differential equation \eqref{e: jetdiffeq}.  Similarly, the antiparallel orientation $\theta = \alpha \pm \pi$ is an unstable equilibium solution of the trajectories, which we likewise characterize as a repulsor of the differential equation \eqref{e: jetdiffeq}.  This attractor / repulsor character of the jet drift with respect to the fluid direction is also apparent even before solving the differential equation \eqref{e: jetdiffeq} from the sign of the moments shown in Fig.~\ref{f:MA1}.  For a negative angle (in this case, an angle clockwise from the fluid flow direction), the jet receives transverse momentum that rotates its angle counterclockwise, and for a positive angle (one that is counterclockwise from the fluid flow direction), it is rotated clockwise. In other words, the jet receives transverse momentum that tends to align it toward the fluid velocity, leading to genuine ``jet drift'' in the direction of the fluid.

To further explore the attractor and repulsor behavior of the trajectories in the neighborhood of $\theta = 0$ and $\theta = \pm \pi$, we can expand the solution \eqref{e: jetdiffeqsol} to obtain the asymptotic behavior close to either the attractor or repulsor.  For a small positive angle $\theta(t) = \epsilon (t)$ from the +x-axis such that $\epsilon\ll1$, we expand the solution \eqref{e: jetdiffeqsol} to the first nonvanishing order in $\epsilon$ and obtain an exponentially converging trajectory
\begin{align} \label{e: expdecay}
    \epsilon (t) \approx \epsilon_{0} \: e^{- t \, / \, \tau_d}
\end{align}
with $\epsilon (0) = \epsilon_0$ and the time constant of the exponential decay given by
\begin{align} \label{e: decaytimeconst}
    \tau_d=\frac{\frac{1}{u}-1}{\frac{3}{\lambda}\frac{\mu^2}{E^2}\ln{\frac{E}{\mu}}} \, .
\end{align}
The same can be repeated again but for the repulsor behavior in the vicinity of $\theta = \pi$. If we let $\theta (t) =\pi - \epsilon(t)$, once again with $0<\epsilon\ll1$, we find that $\epsilon$, now representing the deviation from the repulsor angle $\theta=\pi$, shows exponential growth with respect to time:
\begin{align}   \label{e:expg}
    \epsilon (t) \approx \epsilon_0 \, e^{+ t / \tau_g}
\end{align}
with the time constant of the exponential growth given by
\begin{align}   \label{e:tg}
    \tau_g=\frac{\frac{1}{u}+1}{\frac{3}{\lambda}\frac{\mu^2}{E^2}\ln{\frac{E}{\mu}}} \, .
\end{align}
As with the slopes at $\theta = 0$ and $\theta = \pi$ in \eq{e:sloperatio}, taking the ratio of the time constants \eqref{e: decaytimeconst} and \eqref{e:tg} cancels out all dependence on the medium properties except for the fluid speed:
\begin{align} \label{e:tcratio}
    \frac{\tau_d}{\tau_g}=\frac{1-u}{1+u} \, .
\end{align}
For any nonzero $u$, the time constant $\tau_d$ for the jet angle close to the attractor is always less than the time constant $\tau_g$ for the angle close to the repulsor.  This gives yet another way to obtain the fluid speed from the trajectory of the jet; solving \eq{e:tcratio} for $u$ gives
\begin{align}
    u=\frac{\tau_g-\tau_d}{\tau_g+\tau_d} \, .
\end{align}
Even if the medium is not large enough to see the eventual convergence of the jet onto the fluid flow direction, the early-time behavior \eqref{e:expg} of jets initially aligned against the fluid direction may carry a valuable imprint of the flow characteristics.

%%%%%%%%%%%%%%%%%%%%%%%%%%%%%%%%%%%%%%%%%%%%%%%%%%%%%%%%%%%%%%%%%%%%%%%%%%%
%
\section{Model (In)Dependence: Hard Thermal Loop Potential in Opacity Expansion Formalism}
\label{sec:HTL}
%
%%%%%%%%%%%%%%%%%%%%%%%%%%%%%%%%%%%%%%%%%%%%%%%%%%%%%%%%%%%%%%%%%%%%%%%%%%%

Most of the calculations we perform here are built upon the Gyulassy-Wang (GW) potential and associated calculations which assume heavy (usually static) sources in the medium, so it bears comparison of our results with those that arise from alternative models to determine what level of model dependence exists. Two ends of the model spectrum employed in the field are the somewhat simplistic Gyulassy-Wang model, a perturbative calculation based on heavy scattering centers, and a more complex Hard Thermal Loop (HTL) framework, a perturbative formaslim valid at temperatures high relative to the QCD confinement scale \cite{Armesto_2012}. We compare these two frameworks because they represent opposite ends of the field of models employed in opacity expansion descriptions of jet-medium interactions. While it would also be intriguing to discuss alternative formulations of this effect, perhaps via the BDMPS (Baier, Dokshitzer, Mueller, Peign\'{e}, and Schiff,) \cite{Baier_1997} or higher twist \cite{Wang_2001} frameworks, that discussion lies outside the scope of this paper.

These two models differ significantly in their microscopic details: the simplistic Gyulassy-Wang treatment is built from ingredients like elastic scattering cross sections which resemble zero-temperature QCD and generate contributions which are finite in all diagrams, whereas the dynamic calculations in HTL suffer from infrared divergences at small $q_\PrpXY$ in individual diagrams and only become infrared finite after an account of all diagrams.

Following Ref. \cite{ Djordjevic:2009cr}, we can see that there exists a one-to-one mapping between the first orders in opacity of the static Gyulassy-Wang and dynamic Hard Thermal Loop models. Neglecting the possibility of a nonperturbative magnetic mass, moving between the formalisms can be accomplished simply by the following replacements of the mean free paths $\lambda$ and effective scattering cross sections $\hat\sigma$:
\begin{subequations}    \label{GWtoHTL}
\begin{align} 
    \rho \: \sigma_{0\,GW} = \frac{1}{\lambda_{GW}} & 
    \qquad \Longleftrightarrow \qquad 
    \frac{1}{\lambda_{HTL}} = 3 \alpha_s T = \frac{1}{c(n_f)\, \lambda_{GW}}  \: ,
    \\
    \frac{\mu^2}{\pi (q_\PrpXY^2 + \mu^2)^2} = \hat{\sigma}_{GW}(q_\PrpXY^2) & 
    \qquad \Longleftrightarrow \qquad 
    \hat{\sigma}_{HTL}(q_\PrpXY^2) = \frac{\mu^2}{\pi q_\PrpXY^2 (q_\PrpXY^2 + \mu^2)} \: ,
\end{align}
\end{subequations}
where the conversion factor $c(n_f)$ is a particular function of the number $n_f$ of light flavors.  A typical value for be $n_f=2.5$ is $c(2.5)=0.84$; see Ref.~\cite{Djordjevic:2009cr} for a full treatment of this function and the conversion mapping.

As clearly seen in Eqs.~\eqref{GWtoHTL}, both GW and HTL potentials behave identically in the UV limit $q_\PrpXY^2 \gg \mu^2$.  This agreement is mandatory in order to describe the short-distance, high-momentum perturbative tail of the gluon distribution from the medium sources.  Where the two models differ significantly is in their treatment of the IR limit $q_\PrpXY^2 \lesssim \mu^2$, with $\hat\sigma_{GW}$ remaining finite down to $q_\PrpXY^2 = 0$ and $\hat\sigma_{HTL}$ diverging logarithmically as $q_\PrpXY^2 \rightarrow 0$.  These two infrared behaviors can be thought of as limiting cases (perhaps approximately in the case of GW\footnote{Clearly the numerator of \eq{e:HTLplus} vanishes if $\mu_M^2 = \mu^2$, so one does not exactly reproduce GW as a valid limit of \eq{e:HTLplus}.  However, if $\mu_M^2 = (1 - \epsilon) \mu^2$ and one works to lowest nontrivial order in $\epsilon \ll 1$, then $\hat\sigma_{HTL + \mu_M} \approx \epsilon \: \hat\sigma_{GW}$.  Although the absolute normalization still differs from GW, this limit does have the same IR behavior $\sim (q_\PrpXY^2 + \mu^2)^{-2}$ as GW.}) of the more general case of HTL including a nonperturbative magnetic mass $\mu_M$ \cite{Djordjevic_2012}:
\begin{align}   \label{e:HTLplus}
    \hat\sigma_{HTL + \mu_M} = \frac{1}{\pi} \, 
    \frac{\mu^2 - \mu_M^2}{(q_\PrpXY^2 + \mu_M^2)(q_\PrpXY^2 + \mu^2)} \: ,
\end{align}
with the strict HTL result corresponding to $\mu_M = 0$ and the IR behavior of GW arising in the limit $\mu_M^2 \approx \mu^2$.  In this sense, the GW and HTL potentials represent the extreme endpoints of the infrared behavior captured by the general expression \eqref{e:HTLplus}, along with the universal ultraviolet behavior of perturbative QCD.

\begin{figure}
    \centering
    \includegraphics[width=0.49\textwidth]{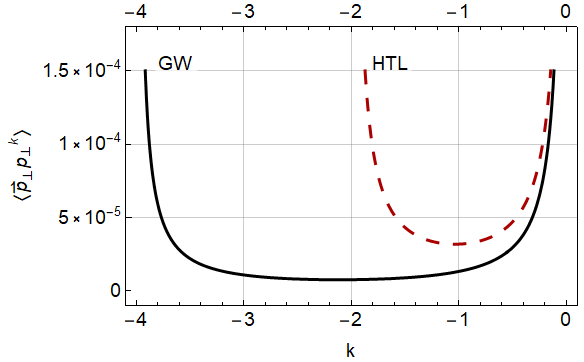}
    \includegraphics[width=0.49\textwidth]{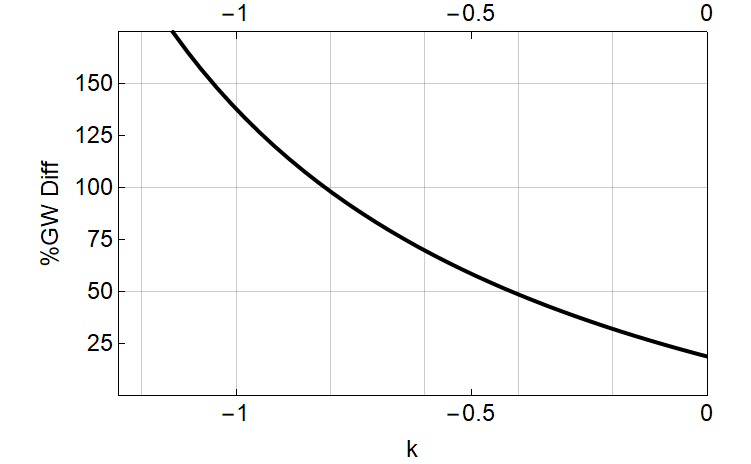}
    \caption{Left: Plot of GW, solid (black), and HTL, dashed (Red), Constant Slab Moments vs $k$, setting $\mu = 1$, $u_\PrpXY = 0.7$, $u_\parallel = 0.1$, $E = 10^6$ GeV, $c(n_f)=0.84$, and our opacity to 1. Right: Plot of \% GW Moment model difference as a function of k.}
    \label{f:HTLvsGW}
\end{figure}

We construct a new expression for the jet $\tvec{p}$ distribution \eqref{e:gen3}, first inserting our potential explicitly:
\begin{align}
    \hat\sigma_{HTL}(q_\PrpXY^2, t) &= \frac{1}{\pi} \: \frac{\mu^2 (t)}{q_\PrpXY^2 (q_\PrpXY^2 + \mu^2(t))^2} & \rightarrow &&
    \frac{1}{\hat{\sigma} (q_\PrpXY^2, t)} \frac{\partial \hat\sigma}{\partial q_\PrpXY^2} &= 
    -\frac{1}{\mu ^2 +q_{\PrpXY}^2} -\frac{1}{q_\PrpXY^2}\: ,
\end{align}
With energy function from \eqref{e:GWsub} and using the prescription \eqref{GWtoHTL} to convert from GW to HTL, we find:
\begin{align}  
    \frac{dN^{(1)}}{d^2 p_\PrpXY} 
    &=
    \int \frac{dt}{c(n_f)\,\lambda(t)} \: \left( \frac{1}{\pi} \: \frac{\mu^2 (t)}{q_\PrpXY^2 (q_\PrpXY^2 + \mu^2(t))^2} \right) \:
    \bigg[ 1 
    + \frac{\tvec{u} (t) \cdot \tvec{p}}{1-u_\parallel (t)} \: \frac{1}{E} \:
    \bigg(
    \frac{q_{\PrpXY}^2}{\mu ^2 +q_{\PrpXY}^2} 
    + 5
    \bigg)
    \bigg]
\end{align}
Where again, we have our unit symmetric term and an asymmetric skew term. We calculate our odd moment as in \eqref{e:vecmoment1}, again noting that our symmetric term integrates to zero, and use our same $\xi$ substitution, thus leading to a modified expression for the transverse momentum moments (c.f. \eq{e:vecmoment2}):
\begin{align}
    \langle \vec{p}_\PrpXY p_\PrpXY^k \rangle_{HTL} & =  \int d^2 p_\PrpXY \, (\tvec{p} \, p_\PrpXY^k) \, \frac{dN^{(1)}}{d^2 p_\PrpXY} \notag \\
    & = \frac{1}{c(n_f)\,E}\int \frac{dt}{\lambda(t)} \frac{\vec{u}_\PrpXY(t)}{1-u_\parallel(t)}  \bigg( \int d\vec{q_\PrpXY} \frac{1}{2 \pi} \left( 5\frac{\mu ^2 q_\PrpXY^k}{ \left(\mu ^2+q_\PrpXY^2\right)}+\frac{\mu
    ^2 q_\PrpXY^{k+2}}{ \left(\mu ^2+q_\PrpXY^2\right){}^2}\right)\bigg) \notag \\
    &=  \frac{I_{HTL}(k)}{c(n_f) E}\int \frac{dt}{\lambda(t)} \frac{\vec{u}_\PrpXY(t)}{1-u_\parallel(t)} \mu^{k+2} 
\end{align}
This makes explicit our relationship with the Gyulassy-Wang moments \eq{e:vecmoment2}. Our line integral through the medium is identical, but we have a slight difference in prefactors. We also see our first marked departure from the simplicity of our Gyulassy-Wang setup: for any $k<0$ we have an unshielded infrared singularity in our first $q_\PrpXY$ integral term. We find, nonetheless, convergence for integration over all transverse momentum space for the moments $k \in (-2,0)$, making our new integral prefactor:
\begin{align} \label{e:HTLI(k)}
    I_{HTL}(k) & \equiv  \int d\xi \, \xi^{k/2} \, \frac{3\xi + 5/2}{(\xi + 1)^3}\frac{\xi+1}{\xi} = \bigg(-\frac{1}{2} \pi ^2 (k+12) \csc \left(\frac{\pi  k}{2}\right) \bigg)
\end{align}
Comparing the two moments explicitly, we note that in the shared region of convergence $k\in(-2,0)$ the HTL to GW ratio is inverse linear in the index $k$:
\begin{subequations}\label{e:HTLGW_Comp}
    \begin{align} \label{e:HTLMom(k)}
        \langle \vec{p}_\PrpXY p_\PrpXY^k \rangle_{HTL}  &= \frac{ 1}{c(n_f) E} \int \frac{dt}{\lambda(t)} \: \frac{\vec{u}_\PrpXY(t)}{1-u_\parallel(t)} \: \mu ^{k+2}(t)   \bigg(-\frac{1}{4} \pi (k+12) \csc \left(\frac{\pi  k}{2}\right)\bigg) \: .\\
        \langle \vec{p}_\PrpXY p_\PrpXY^k \rangle_{GW}  &= \frac{1}{E} \int \frac{dt}{\lambda(t)} \: 
        \frac{\tvec{u} (t)}{1-u_\parallel (t)} \:
        \mu^{k+2} (t) \left( - \frac{\pi}{8} (k+2)(k+12) \, \csc \left(\frac{\pi  k}{2}\right) \right)
    \end{align}
\end{subequations}
\begin{align} \label{e:HTLGW_Rat}
    \Rightarrow \frac{\langle \vec{p}_\PrpXY p_\PrpXY^k \rangle_{HTL}}{\langle \vec{p}_\PrpXY p_\PrpXY^k \rangle_{GW}} & = \frac{2}{c(n_f)(k+2)} \approx \frac{2}{0.84(k+2)} .
\end{align}
We illustrate the relationship between the Hard Thermal Loop moment \eqref{e:HTLMom(k)} and the corresponding Gyulassy-Wang moment \eqref{e:vecmoment2} for the case of a constant slab medium in Fig. \ref{f:HTLvsGW}.
%
%--------------------------------------------------------------------------
%
%
%--------------------------------------------------------------------------
Encouragingly, we see that in the ultraviolet range selected by the $k \to 0$ moment, we have complete convergence of the two models, as expected. From the ratio \eqref{e:HTLGW_Ratk0} of the moments in the limit as $k\to0$, we see the relationship agrees with the mapping framework set forth by Djordjevic \cite{Djordjevic:2009cr}, demonstrating their universality more explicitly:
\begin{align} \label{e:HTLGW_Ratk0}
\frac{\langle \vec{p}_\PrpXY \rangle_{GW}}{\langle \vec{p}_\PrpXY \rangle_{HTL}} & = c(n_f) \approx 0.84 \: .
\end{align}
From this short exercise, we conclude that the physics of jet drift is universal for the $k=0$ moment in any model, so that the conclusions about the deflection angle $\langle \Delta \theta \rangle$ and mean jet trajectories in Appendix ~\ref{sec:trajectory} are quite general and model independent.  For any moments in the range $k\in(-2,0)$, the range of models encompassed by HTL and GW are convergent, possess the same qualitative behavior, and differ only in the numerical prefactor shown in the right panel of Fig.~\ref{f:HTLvsGW}.  Moments for $k\in(-4,-2)$ are more heavily weighted toward the infrared regime where the models differ; while moments in this range may be convergent in some models, conclusions about these moments should be considered model-dependent. While this limits our range of safe observables to study the physics of jet drift, it also raises the potential to use these infrared-weighted moments for experimental model discrimination.

Finally, as we allude to in our outlook in Sec.~\ref{sec:concl}, the $k=-2$ moment has a particularly important role beyond jet acoplanarities due to its emergence in the calculation of asymmetric soft gluon radiation. While in the Gyulassy-Wang model this moment is finite, in the Hard Thermal Loop calculation as we have presented it it is logarithmically divergent. We need to regulate the integral \eqref{e:HTLMom(k)} with an infrared cutoff to extract the leading-logarithmic behavior. This regulator has a good candidate physical interpretation: the magnetic mass $\mu_M$ from \eq{e:HTLplus} which acts to screen the IR behavior at a scale below the Debye mass $\mu$. Indeed, by substituting the general form \eq{e:HTLplus} into \eq{e:gen3} we come to an explicit result \eqref{e:HTLk-2} in the leading logarithmic approximation, regulated by the magnetic mass.
\begin{align}   \label{e:HTLk-2}
\langle \vec{p}_\PrpXY / p_\PrpXY^2 \rangle_{HTL} & = \frac{ 1}{ c(n_f) E}  \int \frac{dt}{\lambda} \frac{\vec{u}_\PrpXY(t)}{1-u_z(t)}  \bigg(5 \log \left(\frac{\mu }{\mu_M}\right)\bigg) \: .
\end{align}
Thus we conclude that the $k = - 2$ moment $\langle \vec{p}_\PrpXY / p_\PrpXY^2 \rangle$ is still logarithmically controllable even in the case of strict HTL with an infrared regulator set by the nonperturbative scale of the magnetic mass $\mu_M$.  By controlling the endpoint behavior in the UV at $k = 0$ and in the IR at $k = -2$ in this way, we can safely draw model-independent conclusions in between these limits about the physics of jet drift and its relation to other observables, up to an overall normalization factor.  For this reason, we argue that the qualitative physics of jet drift we enumerate throughout this paper is quite general and not limited to a particular microscopic model of the medium.

%\bibliographystyle{bibstyle}
%\bibliography{Refs}

\end{document}